\documentclass[11pt,preprint,graphicx]{aastex}
\usepackage{graphicx}
\usepackage[labelfont={sf,bf}, margin=1cm]{caption}

\def\etal{\it et al. \rm }

\begin{document} 

\title{Stellar Populations and the Star Formation Histories of LSB Galaxies:
I. Optical and H$\alpha$ Imaging}

\author{James Schombert}
\affil{Department of Physics, University of Oregon, Eugene, OR 97403;
jschombe@uoregon.edu}

\author{Tamela Maciel}
\affil{Department of Physics, University of Oregon, Eugene, OR 97403;
tmaciel@uoregon.edu}

\author{Stacy McGaugh}
\affil{Department of Astronomy, University of Maryland, College Park, MD 20742;
ssm@astro.umd.edu}

\begin{abstract}

\noindent This paper presents optical and H$\alpha$ imaging for a large sample of LSB
galaxies selected from the PSS-II catalogs (Schombert \etal 1992).  As noted
in previous work, LSB galaxies span a range of luminosities ($-10 > M_V >
-20$) and sizes (0.3 kpc $< R_{V25} <$ 10 kpc), although they are consistent
in their irregular morphology.  Their H$\alpha$ luminosities (L(H$\alpha$)
range from 10$^{36}$ to 10$^{41}$ ergs s$^{-1}$ (corresponding to a range
in star formation, using canonical prescriptions, from 10$^{-5}$ to 1
M$_{\sun}$ yr$^{-1}$).  Although their optical colors are at the extreme
blue edge for galaxies, they are similar to the colors of dwarf galaxies
(van Zee 2001) and gas-rich irregulars (Hunter \& Elmegreen 2006).
However, their star formation rates per unit stellar mass are a factor of
ten less than other galaxies of the same baryonic mass, indicating that
they are not simply quiescent versions of more active star forming
galaxies.  This paper presents the data, reduction techniques and new
philosophy of data storage and presentation.  Later papers in this series
will explore the stellar population and star formation history of LSB
galaxies using this dataset.

\end{abstract}

\section{Introduction}

The key to understanding the evolution of late-type galaxy systems is their
star formation history.  For while a majority of their stellar mass
originates from the epoch of galaxy formation, their current visual
appearance is driven by star formation over the last Gyr (Gallagher \etal
1984).  Thus, studies of the characteristics of star forming galaxies is a
glimpse into the process of star formation and, thus, a window into the
conditions that played a role during the galaxy formation era.

Late-type galaxies come in a range of morphological appearances and their
study has been, for decades, been dominated by the extremely bright,
actively star forming examples such as NGC 4449 (Huchra \etal 1983).  The
advent of newer all-sky surveys in the 1980's/90's demonstrated the
importance of low surface brightness (LSB) galaxies to the galaxy
population, and opened up a wider range of irregular late-type galaxies for
study.  Although initial suggestions were that LSB galaxies dominate the
total galaxy population of the Universe over their higher surface
brightness (HSB) cousins, it was later found to be untrue (Rosenbaum \&
Bomans 2004, Hayward, Irwin \& Bergman 2005).  Nonetheless, LSB galaxies
offer a new avenue for the study of galaxy evolution, having low stellar
densities and recent star formation rates.  Their study, as a class of
galaxies, has merit for stellar population work.

Star formation in galaxies falls into three crude categories: 1)
hyper-efficient, typically associated with tidal or merger events (Whitmore
\etal 2005), 2) normal, associated spiral galaxies under density wave or
stochastic processes (Young \etal 1996) and 3) bursts followed by long
quiescent periods, typically associated with dwarf galaxies (Schombert
\etal 2001).  LSB galaxies, almost by definition, are found in the third
category and are assumed to have had below average star formation rates for
their entire lifetimes to explain their star-forming blue colors, yet low
stellar densities.

\begin{figure}[!t]
\centering
\includegraphics[scale=0.7,angle=0]{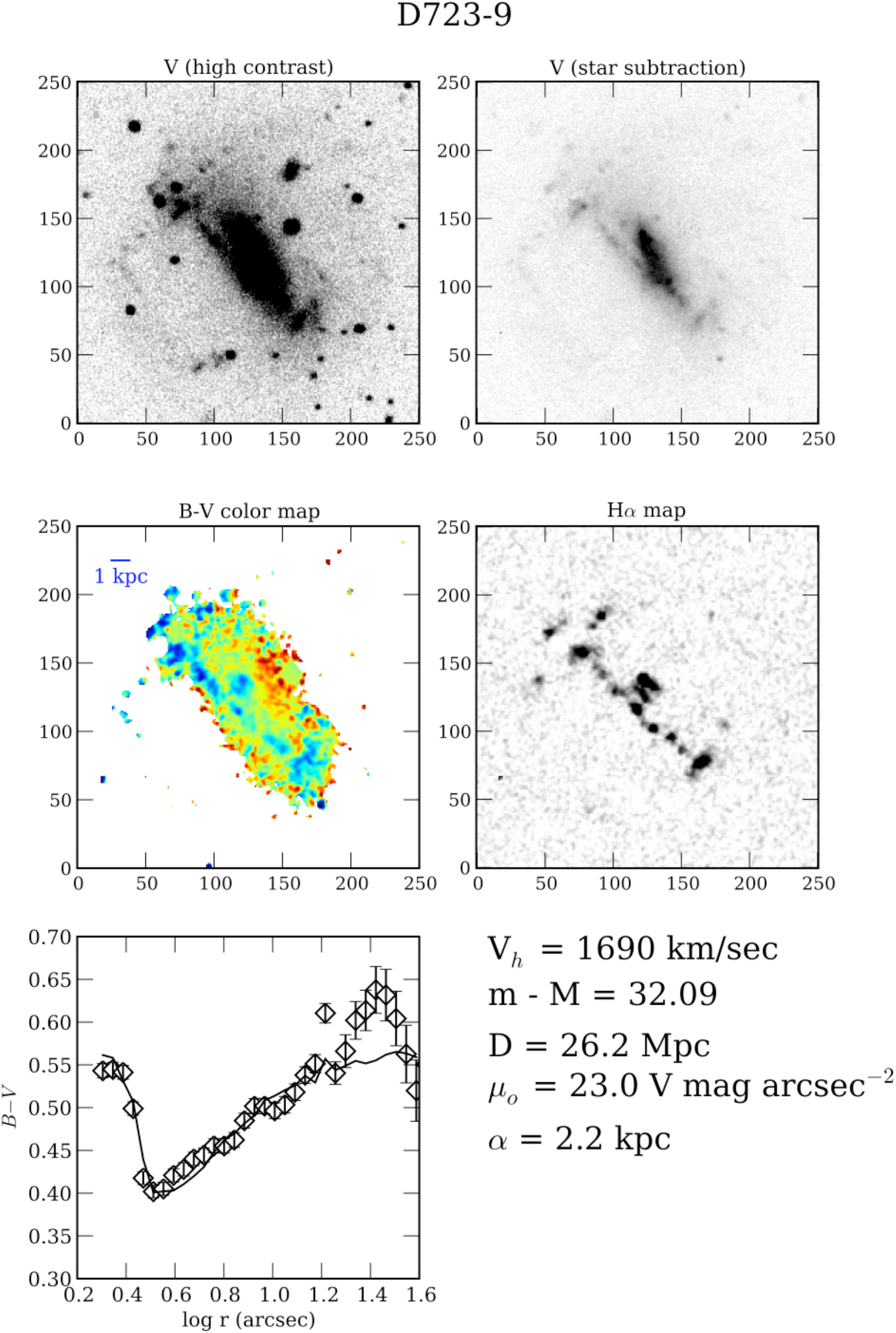}
\caption{\small An example visual summary chart for galaxy D723-9.  The top
left panel is a high contrast $V$ image (to emphasize LSB features),
axes are in pixel units, north at top, east to left.  The
top right panel is a low contrast $V$ image with foreground stars removed.
The middle left panel is the $B-V$ color map (blue is $B-V=0$, red is
$B-V=1$).  The middle right panel is the H$\alpha$ emission.  The lower
left panel is the $B-V$ color profile (open symbols are differential color,
solid line is integrated color).  Redshift, distance, central surface brightness
and scalelength ($\alpha$) are also listed.
All the galaxy's charts can
be found at http://abyss.uoregon.edu/$\sim$js/lsb.
}
\end{figure}

This project is focused on obtaining optical and H$\alpha$ imaging of a
sample of LSB galaxies with known HI properties.  For knowledge of the blue
optical colors, H$\alpha$ emission and HI gas mass provides constraints on
the dominant stellar population for the last few Gyrs, the current star
formation rate and the potential total star formation rate (i.e., the
amount of material available to produce stars).  Integrated values speak to
the global history of star formation, but, in addition, spatial color and
H$\alpha$ information can distinguish between episodic bursts of star
formation versus quasi-continuous modes (Gerola, Seiden \& Schulmann 1980,
van Zee 2001).

\begin{figure}[!t]
\centering
\includegraphics[scale=0.7,angle=0]{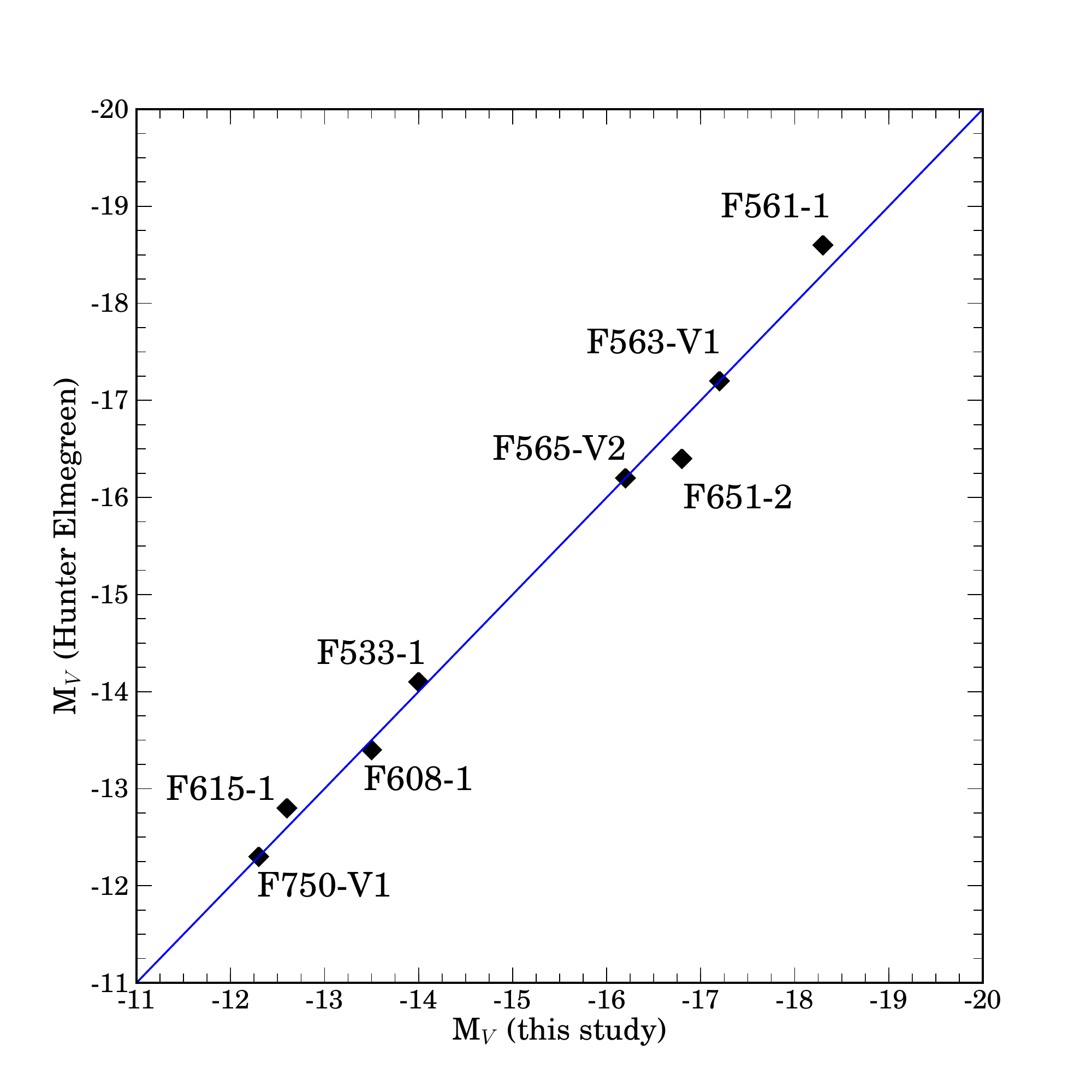}
\caption{\small A comparison between $V$ magnitudes determined by Hunter \&
Elmegreen and this study.  The one-to-one correspondence line is shown in
blue.  The mean difference is only 0.01 mags.  The larger disagreement
for the brightest galaxies (F561-1 and F651-2) is due to the our technique
of replacing stars with galaxy light.
}
\end{figure}

In this paper, we present color and H$\alpha$ imaging for a large
sample of LSB galaxies, not restricted by size or total luminosity.  As
these galaxies also have aperture HI measurements, this sample will
comprise one of the largest sets of LSB galaxy data to date with complete
stellar and gas observations.  We will present the techniques, analysis and
reduced data for $B$, $V$ and H$\alpha$ imaging and comparison to other
samples, but will reserve interpretation and modeling for later papers.  We
will also present the data in a new format so that interested readers can
extract the reduced values as well as the scripts that were used to convert
the apparent values to astronomically meaningful values.  This brings a new
level of transparency to datasets where the individual can not only follow
the corrections and transformations applied to the raw data, but substitute
their own values and seamlessly produce their own final datasets.

\section{Sample Selection and Observations} 

Our sample of LSB galaxies was primarily selected from the catalog of
Schombert \etal (1992) and Schombert, Pildis \& Eder (1997).  Those project's
goals were to observe a large number of LSB dwarf galaxies to test biased
galaxy formation scenarios.  The resulting sample was by no means
restricted to dwarf galaxies (LSB appearance does not correlate with
galaxy mass, Pildis, Schombert \& Eder 1997).  A number of LSB galaxies
from the original PSS-II catalog (Schombert \& Bothun 1988) were also
included.  The basic criteria for the PSS-II catalog was to find objects
that simply do not fit into the normal Hubble sequence, such as spirals or
HSB irregulars, and were LSB in character (this visual selection was later
determined to be approximately $\mu_c >$ 23 $V$ mag arcsec$^{-2}$).

Target selection for follow-up optical and H$\alpha$ imaging used a
combination of object morphology, redshift and position in the sky.
Astrophysically, we wished to sample a range of sizes, luminosities and HI
masses.  Filter constraints for H$\alpha$ imaging restricted the redshift
range to less than 8,000 km/sec.  All the objects are between $-$10 and
$+$30 declination (the Arecibo window).  Observing schedules restricted the
right ascension window to between 23H and 4H in the fall, 8H and 13H in the
spring.  An effort was made to avoid galaxies with clear spiral patterns,
as they have been studied by others (Kennicutt \etal 2008).

The final sample is given in Tables 1 and 2.  The columns in Table 1 are as
follows: 1) galaxy name, 2) run number (K, for KPNO, plus MMYY), 3)
distance in Mpc (taken from NED), 4) absolute $V$ magnitude, 5) central
surface brightness from exponential fit, 6) scalelength in kpc, also from an
exponential fit, 7) mean $B-V$ color, 8) mean $V-I$ color and 9) axial
ratio at the 25 $V$ mag arcsecs$^{-2}$ isophote.  The columns in Table 2
are as follows: 1) galaxy name, 2) H$\alpha$ flux, 3) H$\alpha$ luminosity,
4) stellar mass, following the prescription of Schombert \etal (2000), 5)
HI gas mass, again from Schombert \etal (2000), 6) baryon mass (stellar
plus gas), 7) the gas fraction and the birthrate function (SFR/$M_*$).

We imaged 59 LSB dwarfs and disks plus the well-studied dwarfs DDO154 and
DDO168 for comparison.  The spiral galaxy UGC 128 was also imaged as a
control HSB disk galaxy.  Selection by morphology for the original catalog
did not exclude luminous or large galaxies, however, selection for galaxies
with redshifts less than 8,000 km/sec (to fit within the H$\alpha$ filter
set) eliminates the high redshift Malin cousins.  We note that a late-type
morphological criteria does select for galaxies with high HI/L ratios
(Schombert, Pildis \& Eder 1997).  The final sample ranges from $M_V =
-$19.5 to $-$12.5 and from 0.1 to 5 kpc in scale length.

The imaging for this project was obtained during three runs (two spring:
K0308 \& K0309, one fall: K1007) on the KPNO 2.1m.  A series of narrow band
filters were used to acquire ON/OFF H$\alpha$ frames.  The KPNO H$\alpha$
filters used were 1391, 1494, 1563, 1564, 1565 and 1566.  These filters are
70\AA\ wide centered at 6620, 6658, 6573, 6618, 6653 and 6709\AA\
respectfully.  This gave us velocity coverage from 0 km/sec to 8,000
km/sec, which contains 80\% of the Schombert LSB catalog.  The ON filter
was selected to match the galaxy velocity, the OFF filter was nearest
filter either above or below in velocity space.

In addition, deep $B$ and $V$ frames were obtained for every galaxy in the
sample during the same observing runs in order to perform two color surface
photometry ($B-V$).   Typical exposure times were 3 sets of 150 sec $V$
frames, 3 sets of 300 sec $B$ frames and 6 pairs of 600 sec H$\alpha$
ON/OFF frames (i.e. a total of one hour ON and one hour OFF).  The plate
scale is 0.61 arcsecs per pixel for a field of view of 10.4 arcmins.  Note
that 23 of these galaxies were also imaged in $V-I$ for a previous project
(Pildis, Schombert \& Eder 1997), this was done on a different detector and
telescope and requires some processing to combine with the current sample.

\section{Data Reduction}

Data reduction followed the standard procedures for low readout noise
CCD's.  Dark subtraction used overscan regions.  Image flattening used dome
flats and no large scale features are seen to the 0.1\% level over the
inner 80\% of each frame.  Calibration used Landolt (1992) standards for
$B$ and $V$, and Stone (1996) standards for the H$\alpha$ filters.
Standard KPNO airmass corrections were applied to all frames, although the
greatest airmass observed was only 1.18.  Multiple frames were registered
using internal stars then clipped-summed to eliminate cosmic rays.

The $V$ frames were then registered, cleaned, then summed for ellipse and
surface brightness fitting.  Sky values were first determined by visually 
assigning sky boxes to regions free from other objects.  These sky boxes
were recorded and used to determine sky in the $B$ and H$\alpha$ frames.
Sky boxes are the most useful method of determining the true sky value for
LSB galaxies as they register any large scale variations across the region
around the galaxy as well as provide a measure of the random error
($\sigma$ within each box) and a measure of the error on the mean value
(the $\sigma$ between the means of each box).  The error in the sky
dominates most of the photometric values for LSB galaxies.  The typical
errors in the mean sky value were between 0.1 and 0.2\%.  This corresponds
to a 1$\sigma$ sky error of 28.3 $V$ mags arcsecs$^{-2}$.

\begin{figure}[!t]
\centering
\includegraphics[scale=0.7,angle=0]{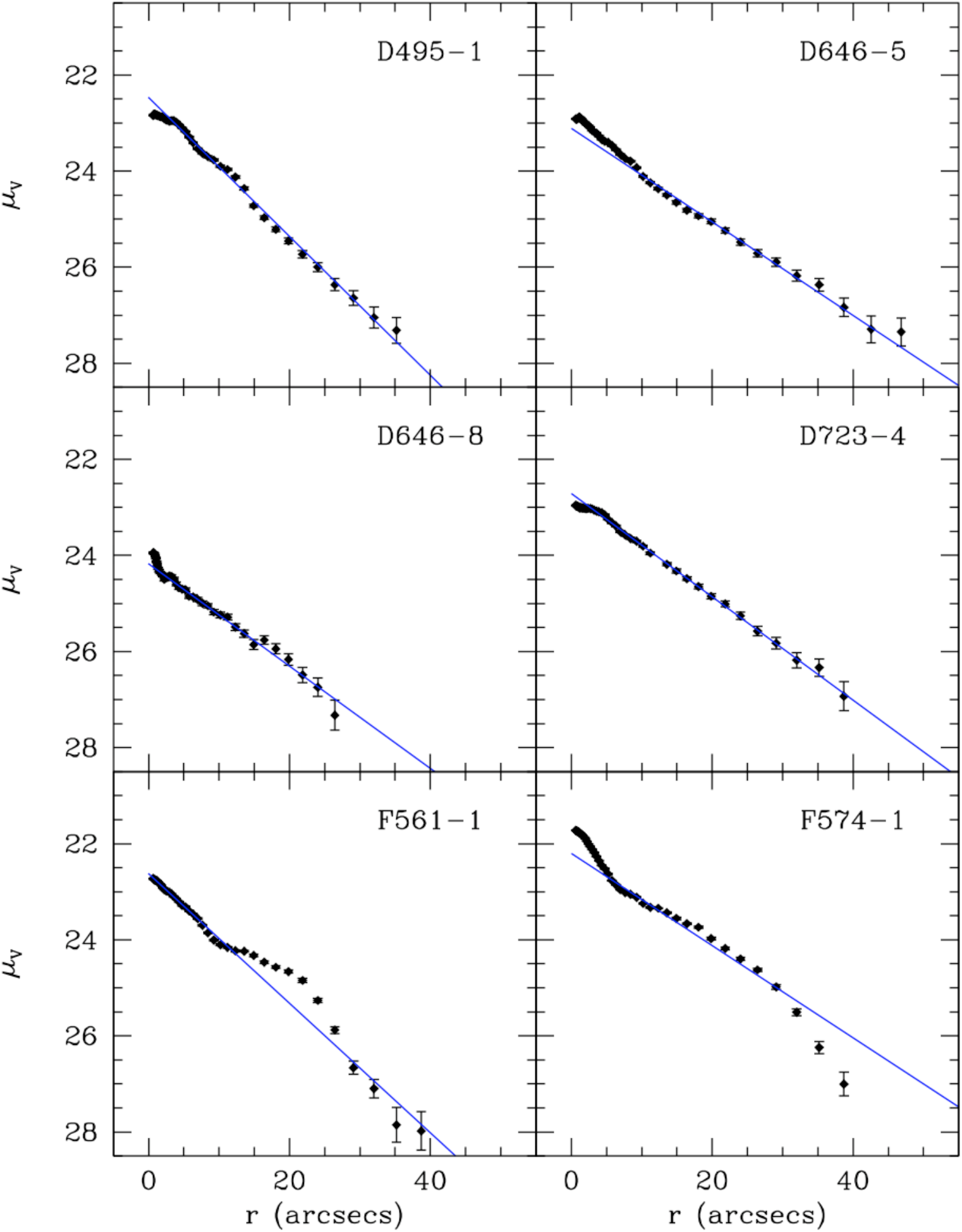}
\caption{\small Surface brightness profiles for six galaxies.  Best fit
exponentials are shown in blue.  Fits are constrained to the outer portions
of the galaxy with the clearest linear behavior.  
}
\end{figure}

The ellipse values for the $V$ frames are used to define apertures for
colors and H$\alpha$ values.  Cleaning was an automatic process of the
ellipse fitting routines (see ARCHANGEL, Schombert 2007).  However, in a
few cases of nearby bright stars or embedded stars, these objects were
manually cleaned.  Cleaned areas were re-filled using intensity values
based on the averaged ellipse that passed through each pixel.  This was
only significant for cleaned stars with the main body of the galaxy and was
never more than 4\% the total luminosity of the galaxy.

Isophotal analysis begins with fitting ellipses to the cleaned image.
Fitting. a best ellipse to a set intensity values in a 2D image is a
relatively straight forward technique that has been pioneered by Cawson
\etal (1987) and refined by Jedrzejewski (1987) (see also an excellent
review by Milvang-Jensen \& Jorgensen 1999).  The core routine from these
techniques (PROF) was eventually adopted by STSDAS IRAF (i.e. ELLIPSE).
The primary fitting routine for this project follows the same techniques
(in fact, uses much of the identical FORTRAN code from the original GASP
package of Cawson) with some notable additions.

These codes start at some intermediate distance from the galaxy core with
an estimated x-y center, position angle and eccentricity to sample the
pixel data around the given ellipse.  The variation in intensity values
around the ellipse can be expressed as a Fourier series with small second
order terms.  Then, an iterative least-squares procedure adjusts the
ellipse parameters searching for a best fit, i.e. minimized coefficients.
There are several halting conditions, such as maximum number of iterations
or minimal/extreme change in the coefficients, which then moves the ellipse
outward for another round of iterations.  Once a stopping condition is met
(edge of the frame or sufficiently small change in the isophote intensity),
the routine returns to the start radius and completes the inner portion of
the galaxy.  A best fit ellipse is always found in all the frames, although
for the more irregular galaxies it is clear that an ellipse is a forced
figure onto the isophotes.  We will discuss the merit of this process in
our structure paper.

Final fits were visually inspected for robustness, converted to 1D
surface brightness profiles and fit to an exponential disk.  A small number
of objects had significant bulges, but r$^{1/4}$ fits to the bulges did not
significantly alter the disk fits due to the small bulge size.  Both surface
brightness profiles, aperture magnitudes and color profiles are based on
these fits.  For consistency, the $V$ ellipse fits were applied to the $B$
and H$\alpha$ frames (after registration) to produce aperture luminosities
and colors (i.e. the same pixels are integrated by aperture in all frames).

All the reduced values, single parameters (e.g., total luminosity) and
array values (e.g., surface brightness profiles) are all stored in XML
format and placed at the data website
(http://abyss.uoregon.edu/$\sim$js/lsb).  In addition, the data website
contains the scripts (written in the Python computer language) which are
used to convert raw telescope values into astronomical meaningful
parameters.  These well commented scripts allow the user to follow all
the details for data reduction, rather than attempting to extract the
procedures from the published text.  Many of the calibrating values (e.g.
Galactic extinction, CMB distance) are obtained over the network (e.g.
NED), and those scripts are also found at the website.  In addition, we
compare our values with galaxies in common from other studies.  However,
certain parameters, such as distance, have changed since the original
studies were published.  Thus, this script's procedures contain all the
information to convert other datasets into a common framework for
comparison to our data.

An additional challenge is to visually present the data for a large range
of galaxy sizes and morphology.  For it is the spatial color and intensity
information that will address many of the star formation issues.
Structural information is summarized by surface brightness profiles, which
are displayed for all the galaxies at the data website.

The image information (appearance, H$\alpha$ and color maps) are summarized
in a fashion shown in Figure 1 (with the whole sample found at the data
website).  This visual summary includes two grayscale images (on the left
at high contrast, on the right at low contrast with nearby stars removed),
a two color ($B-V$) map (blue is $B-V=0.0$, red is $B-V=1.0$), a high
contrast H$\alpha$ map, the $B-V$ color profile and list of the galaxy's
structural parameters. 

\begin{figure}[!t]
\centering
\includegraphics[scale=0.7,angle=0]{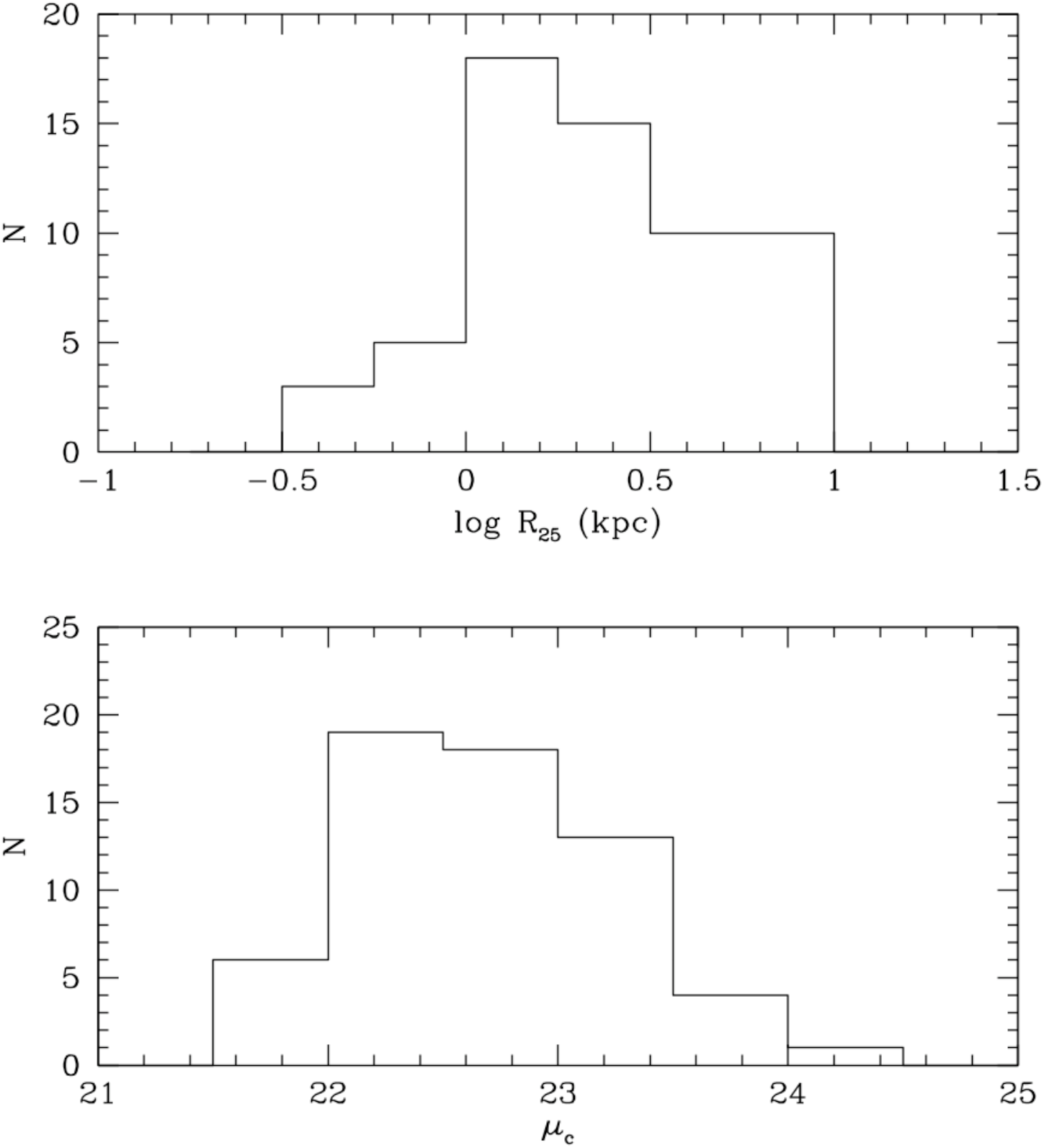}
\caption{\small The range of size and central surface brightness ($V$ mag
arcsecs$^{-2}$) for the
sample.  As demonstrated in early work, not all LSB galaxies are dwarf
galaxies in terms of size (although the LSB giants, e.g. F568-6, were
excluded from the sample due to the restrictions on redshift range).
}
\end{figure}

Note that absolute values are based on distances taken from NED (NASA's
Extragalactic Database) using the concordance model (i.e. NED's
cosmology-corrected distance).  All redshifts were based on 21-cm HI
measurements (Eder \& Schombert 2000).  For nearby objects in the Hunter \&
Elmegreen (2006) and van Zee (2001) samples, redshift independent distances
(e.g. Cepheids) from NED were used where available.

\subsection{Total Magnitudes/Colors}

Total magnitudes and integrated colors used the cleaned frames where the
cleaned areas are re-filled with estimated intensities from the fitted
ellipses.  While this is not perfectly accurate for irregularly shaped
galaxies, the filled regions rarely contributed more than 4\% the total
light of a galaxy.

Total magnitudes were determined using asymptotic fits to the aperture
photometry (see Schombert 2007).  Rather than using curves of growth (which
are ill-defined for LSB galaxies), these fits were made to (1) the raw pixel
summed values, (2) the intensities calculated from the fitted ellipses, (3)
extrapolation of luminosity from exponential fits to the surface
photometry.  For LSB galaxies, a larger fraction of their luminosity is
found in their halo regions compared to HSB galaxies.  Unfortunately,
these outer pixels are also the closest to sky values and suffer from the
highest errors.  Thus, technique (1) frequently fails to converge due to
noise at low levels.  Likewise, technique (3) is error prone due to the
sensitivity of exponential fits to noisy outer isophotes.

We found reasonably stable total luminosity values using technique (2)
extrapolating the isophotal intensities.  Here the apertures are integrated
to a user specified point (typically the point where $L = 80$\% $L_T$),
then the mean intensities from the fitted ellipses are used to sum the
remaining luminosity.  For a majority of the galaxies imaged in this study,
the total $V$ magnitudes converged using this technique with typical
internal errors of 0.06 mags.  They are listed in Table 1, errors are
assigned based on the error in the sky value (which dominates the noise in
LSB images).

\begin{figure}[!t]
\centering
\includegraphics[scale=0.3,angle=0]{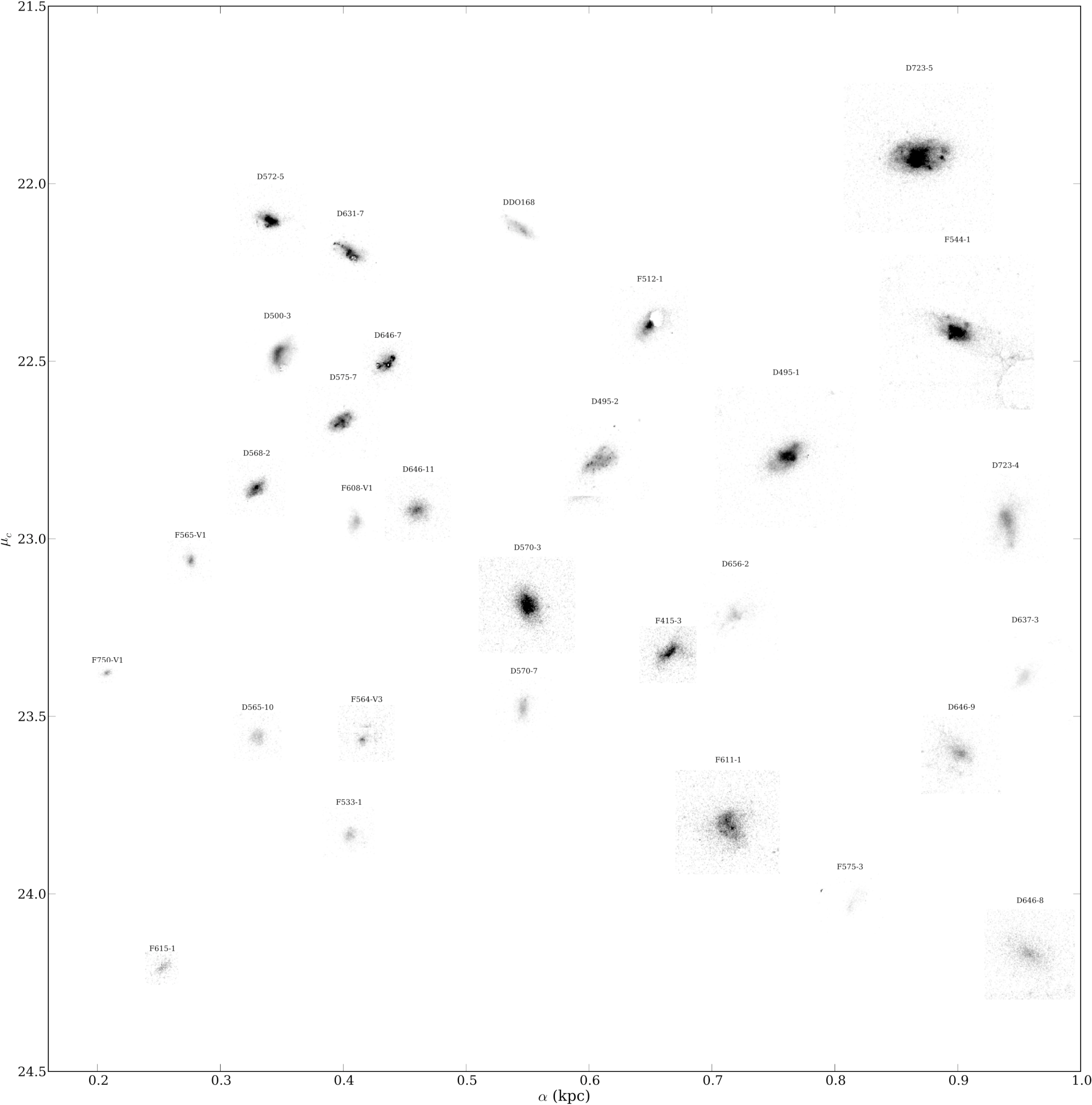}
\caption{\small Galaxy morphology as a function of central surface
brightness ($\mu_c$) and scalelength ($\alpha$).  This panel displays
galaxies with scalelengths less than 1 kpc.  Irregular morphology dominates
LSB galaxies, although smooth morphology is the signature of a gas-poor LSB
and, therefore, would not be included in our sample due to a lack of an HI
redshift.  A larger version of this figure is available at our website.
}
\end{figure}

\begin{figure}[!t]
\centering
\includegraphics[scale=0.32,angle=0]{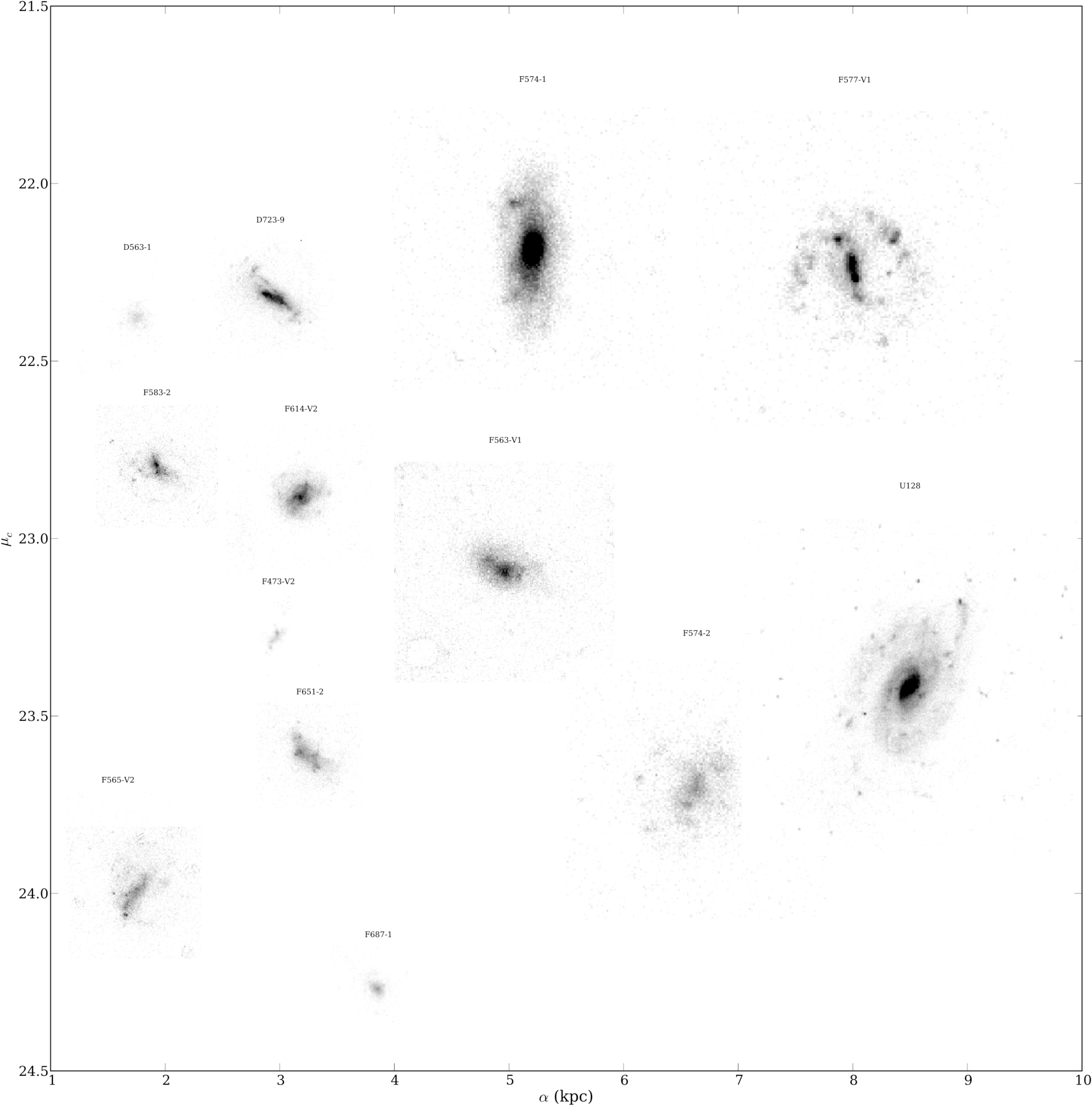}
\caption{\small Galaxy morphology as a function of central surface
brightness ($\mu_c$) and scalelength ($\alpha$).  This panel displays
galaxies with scalelengths greater than 1 kpc.  Flocculate spiral patterns
are evident for the larger LSB galaxies.  A larger version of this figure
is available at our website.
}
\end{figure}

An external check to the integrated magnitudes is offered by comparison to
the LSB sample from Hunter \& Elmegreen (2006).  There are eight objects in
common with that study.  Their identifications and apparent magnitudes
(corrected for galactic extinction) are shown in Figure 2.  
The mean difference is only 0.01 mags, which is well within the formal
errors of 0.06 mags.

Colors are calculated as integrated and differential colors using the $V$
fitted ellipses as apertures.  A large percentage of the LSB galaxies in
this sample have highly variable colors in the spatial sense.  An
integrated color does not capture the whole stellar population picture for
many of these systems.  Often a more accurate global color is one where
the ellipses are used to determine differential color (the color of the
annulus around each ellipse), then these annuli are averaged (weighted by
the surface brightness of the annuli).  It is these weighted colors (out to
the 25 $V$ mag arcsecs$^{-2}$ isophote) that is listed in Table 1 as
$<B-V>$ and $<V-I>$.

The full color information is obtained from pixel-by-pixel two color maps.
These maps are made by registering to the $V$ frame then binning 3x3.  An
example of a spatial map is shown in Figure 1.  It is also possible to plot
spatial color versus surface brightness (i.e. pixel color versus that
pixels surface brightness in $V$).  An example of that type of analysis is
shown in \S3.5.

A subset of the sample had been previously imaged in $V-I$ (Pildis,
Schombert \& Eder 1997).  While those colors are less accurate, their total
values will be compared to the $B-V$ colors in \S3.4 and are found listed
in Table 1.  Spatial $V-I$ color maps are made and re-pixeled to the same
orientation and scale of the newer $B-V$ frames.  This allows for a
comparison of $B-V$ and $V-I$ not only in total colors and color profiles,
put also on a pixel-by-pixel basis.

\subsection{Scale Length ($\alpha$) and Central Surface Brightness ($\mu_o$)}

Surface photometry was extracted from the ellipse fitting using the
standard techniques outlined in Schombert (2007).  Due to their typically
irregular morphology, LSB galaxies are notoriously difficult to simplify
into a 1D light profile.  The procedure used herein is to convert the best
fit ellipses into a surface brightness profile of intensity versus
semi-major axis.  A section of the surface brightness profile is selected
in the outer regions of the galaxy with the most linear appearance.  This
region is then fit to a straight line, interpolating to the core to extract
the central surface brightness ($\mu_o$) and exponential scale length
($\alpha$).  Although the interiors of LSB galaxies are frequently poorly
fit by an exponential profile, the exterior regions are often easily
described by an exponential law.  This is surprising given their irregular
outer isophotes and is probably due to the fact that any high surface
brightness lumps are restricted to their core regions.

A series of example surface brightness fits are found in Figure 3.  These
profiles are a subset of the total sample demonstrating good and poor
(e.g., F561-1) fits, as well as examples where an exponential fit may not
be appropriate.  The full set of surface brightness fits can be found at
our data website.  As can be seen from several of the surface brightness
fits, the fitting technique can lead to a mismatch between the central
surface brightness as described by a fit to the galaxy's outer regions,
versus a central surface brightness that actually represents the luminosity
of the core regions.  To fix this mismatch, the inner five arcsecs are also
fit to extract a true central surface brightness (which we will designate
as $\mu_c$ to distinguish it from $\mu_o$ from exponential fits) except in
cases where there is a clear bulge-like central region (e.g., F574-1).  No
inclination corrections are applied to $\mu_o$ since the intrinsic 3D shape
of LSB galaxies is unknown (some are rotators, others are not, so a thin or
thick disk correction would vary from galaxy to galaxy).

The range of size and central surface brightness is shown in Figure 4.
Here size ($R_{V25}$) is defined by the 25 $V$ mag arcsecs$^{-1}$ isophote
major axis in kpcs.  Central surface brightness is the inner interpolation
to zero radius (in the case of a bulge and disk, interpolation of the
disk).  Sizes range from 0.3 kpc to 10 kpc.  LSB galaxies include both the
smallest of dwarf galaxies up to medium sized disk galaxies.  The central
surface brightnesses are all fainter than the Freeman value (21 $V$ mag
arcsecs$^{-2}$), but do not include the extremely faint surface brightnesses
($mu_o >$ 23 to 24 mags arcsecs$^{-2}$) found in recent digital surveys
(Zhong \etal 2008, Adami \etal 2006), primarily due to the limitations in
the photographic medium used to discover the sample.

To further display the range in size and morphology, Figures 5 and 6
display a greyscale image of selected galaxies (stars removed) with respect
to their central surface brightness and disk scale length.  Irregular
morphology is most common at small scale lengths, although it would be
difficult to predict which galaxies were dwarfs simply from an estimate
based on their appearance.  LSB galaxies with smooth morphologies would
probably be dE's and their lack of HI gas for redshift determination would
exclude them from our samples.  Flocculate spiral patterns are evident at
scale lengths greater than 1 kpc (note, UGC 128 is a HSB counter-example).

\subsection{H$\alpha$ Imaging}

The most observationally intensive portion of this project was obtaining
H$\alpha$ images for the entire sample.  The H$\alpha$ emission for LSB
galaxies is known to be below rates standard for late-type galaxies
(McGaugh, Schombert \& Bothun 1995).  Thus, we anticipated that a majority
of telescope time would be spent on ON/OFF exposures with narrow band
filters.  Indeed, we typically spent 60 mins total on each object in the
sample and were rewarded with a 93\% success rate for H$\alpha$ detection
(all but four of our original 60 targets, see Table 1).

The high detection rate was surprising since, given the low surface
densities and previous H$\alpha$ studies on this sample (McGaugh, Schombert
\& Bothun 1995), there was an expectation that a majority of LSB galaxies
were in quiescent mode with no star formation in the last few Gyrs.
H$\alpha$ surveys of HI-rich galaxies (Meurer \etal 2006) and volume
complete samples (Kennicutt \etal 2008) have high detection rates, thus our
pessimism seemed unwarranted.  We note that the H$\alpha$ fluxes were
extremely low (the mean is a factor of ten less than studies of gas-rich
dwarfs, van Zee 2001), and the four objects with non-detections were the
reddest galaxies in the sample (suggesting an elliptical-like history of an
initial burst, but very little star formation after that).

\begin{figure}[!t]
\centering
\includegraphics[scale=0.7,angle=0]{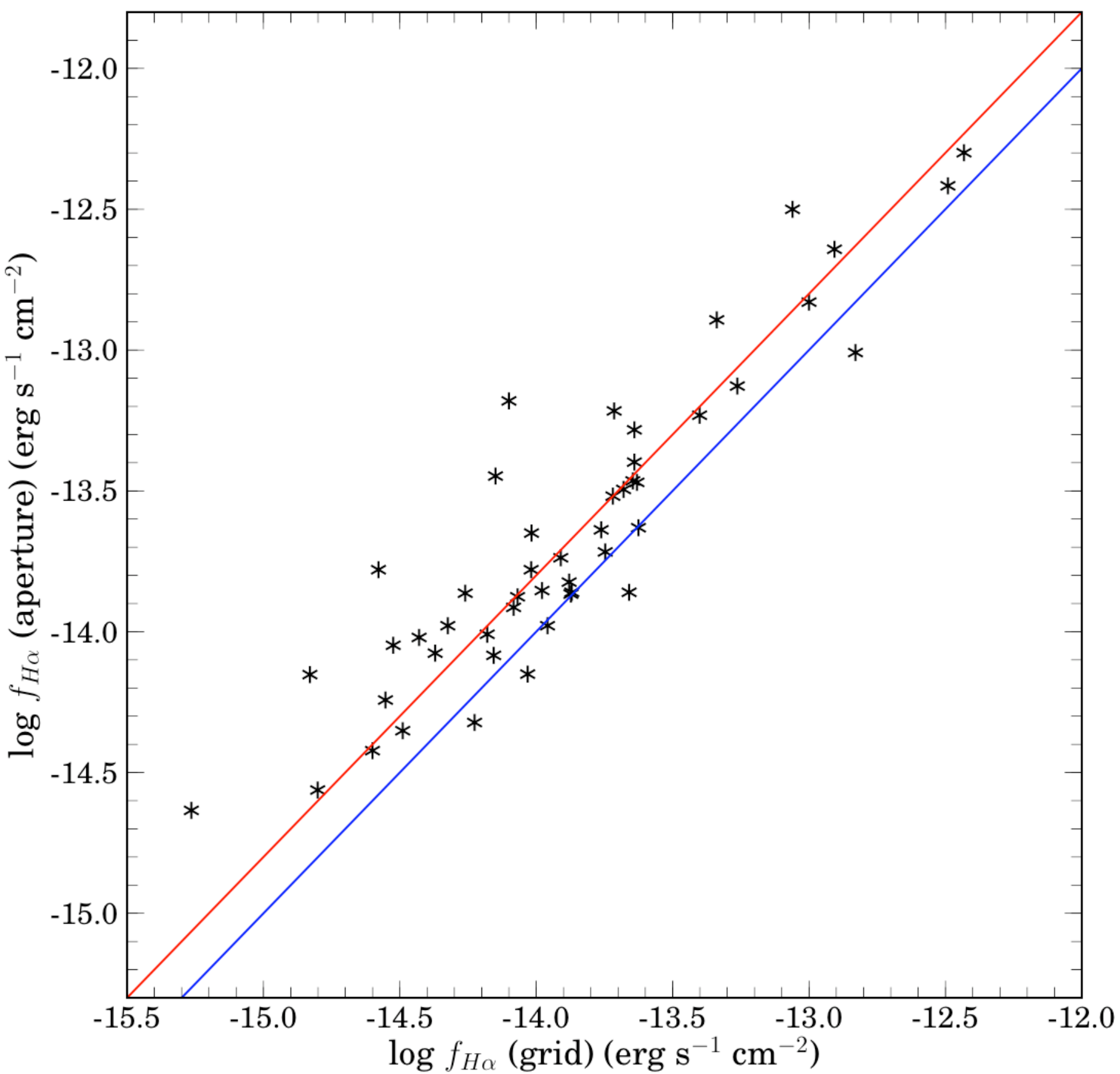}
\caption{\small Comparison of the H$\alpha$ flux determined by grid
photometry of the HII regions versus aperture photometry of the entire
galaxy volume.  The red line is a linear fit to the data, the blue line is
the one-to-one correspondence line.  The aperture values are, on average,
40\% higher than the grid photometry values.  This can be attributed to
diffuse emission presumably spread throughout the galaxy's stellar
population (van Zee 2000).
}
\end{figure}

Our observing technique was fairly standard.  We used the KPNO H$\alpha$
set, which cover the redshift range from zero to 10,000 km/sec with
effective widths of 3,000 km/sec.  Flattening used dome lamps.  All the
frames had noticeable curvature at the edges; however, all the sample
galaxies occupied only the inner 20\% of the chip.  Calibration used Stone
spectrophotometric standards (Stone 1996).

The distribution of H$\alpha$ emission is typically irregular in galaxies,
even for HSB spirals.  The use of elliptical apertures would introduce
needless sky noise, so we adopted a system of mask photometry to measure
the H$\alpha$ flux for each galaxy. For this sample, we use a boxcar
smoothed H$\alpha$ frame to define the initial mask.  Peak
H$\alpha$ regions are located, then those regions are allowed to grow by
20\%.  All the masks are visually inspected and additional areas are added
as needed (usually to capture diffuse emission areas).  This technique has
the advantage of minimizing sky noise and allowing us to directly compare
the $B-V$ colors of the H$\alpha$ pixels.  

There is concern that diffuse H$\alpha$ emission that is not visually
detected would be excluded.  To test for this effect we plot the aperture
values versus the grid photometry values in Figure 7.  The unity line is
marked in blue, a linear fit with a fixed slope of one is shown in red.
The difference between the aperture values and the grid values is 40\%.
This additional H$\alpha$ luminosity is, presumingly, diffuse H$\alpha$
emission that was too low in surface brightness per pixel to be detected by
our growth algorithms.

We note that this 40\% difference is nearly identical to the value for
diffuse emission found by van Zee (2000) of 50\%.  Using the $R_{V25}$
radii, the radius where the surface brightness profile drops to 25 $V$ mag
arcsecs$^{-2}$, as compared to the area of detected H$\alpha$ emission, we
find that visible H$\alpha$ covers a range from 5 to 25\% the total area of
the galaxy, with a mean value for the sample of 10\%.  The diffuse
H$\alpha$ emits from this remaining 90\% of the galaxy volume, presumably
from the bluest regions.  In order to capture all the H$\alpha$ flux for
star formation rate calculations, we have used our grid values (with the
lowest errors) and increased them by 40\% to account for the diffuse
component.

\begin{figure}[!t]
\centering
\includegraphics[scale=0.7,angle=0]{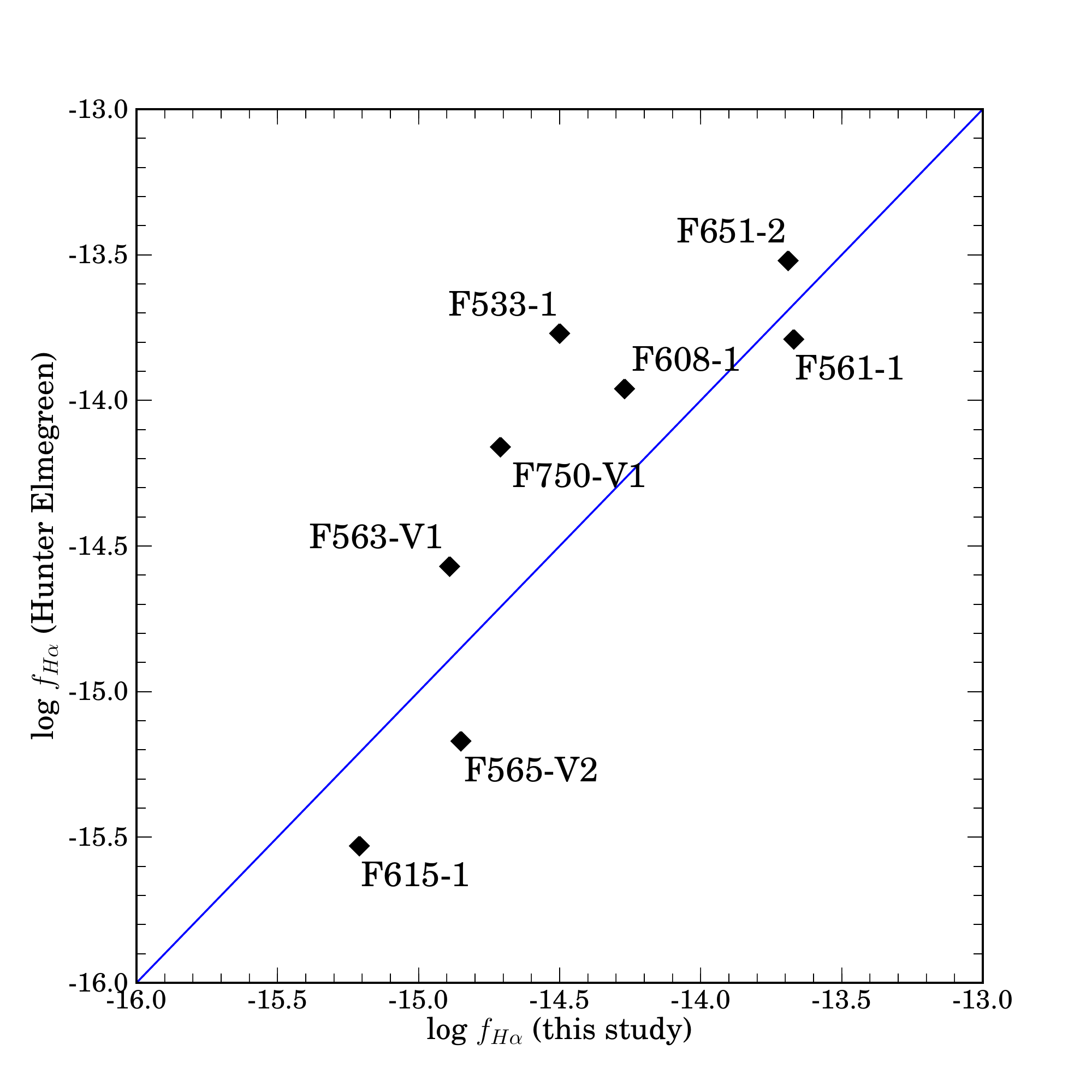}
\caption{\small Comparison between H$\alpha$ fluxes from Hunter \&
Elmegreen and this study.  The blue line delineates a 1-to-1 correspondence.
The average difference is 0.15 in log flux, the standard deviation is 0.4
in log flux.
}
\end{figure}

There are six galaxies in common, in terms of H$\alpha$ fluxes, between our
study and the work of Hunter \& Elmegreen (2004).  The Hunter \& Elmegreen
is plotted against our values in Figure 8.  The average difference is 0.15
in log flux, the standard deviation for the eight galaxies is 0.4.  There
was no correlation between apparent H$\alpha$ flux and the residuals
between the two samples.

Since all our galaxies were imaged with the same number of ON/OFF
exposures, we can estimate, based on the flatness H$\alpha$ frames and
integration times, that our upper limit for H$\alpha$ detection is
approximately 10$^{-16}$ ergs s$^{-1}$ cm$^{-2}$.  Since our exposure times
were similar from galaxy to galaxy, and night sky conditions also were
stable during the run, we consider this upper limit to be constant for this
study.  For the four galaxies without an H$\alpha$ detection, this
corresponds to log $L(H\alpha)$ between 36 and 37, which would be on the
faint side of our $L(H\alpha)$ values.

The total H$\alpha$ fluxes and H$\alpha$ luminosities are listed in Table
2.  The H$\alpha$ fluxes are corrected for an assumed [NII]6583+6548/H$\alpha$ ratio
of 0.1 (Hunter \& Elmegreen 2004) and a factor of 40\% to account for
diffuse emission (see above).  No corrections for dust extinction are made
as LSB galaxies have very few far-IR detections and the appropriate
corrections are unclear.  The resulting fluxes are then converted to
L(H$\alpha$) using the prescription given in van Zee (2001).  A histogram
of L(H$\alpha$) values is shown in Figure 9.   For comparison, the bright
dwarf sample of Hunter \& Elmegreen (2004) and van Zee (2001) are also
shown.  The mean H$\alpha$ luminosity of our sample is about a factor of 3
fainter than Hunter \& Elmegreen, but similar to van Zee's.

\begin{figure}[!t]
\centering
\includegraphics[scale=0.5,angle=0]{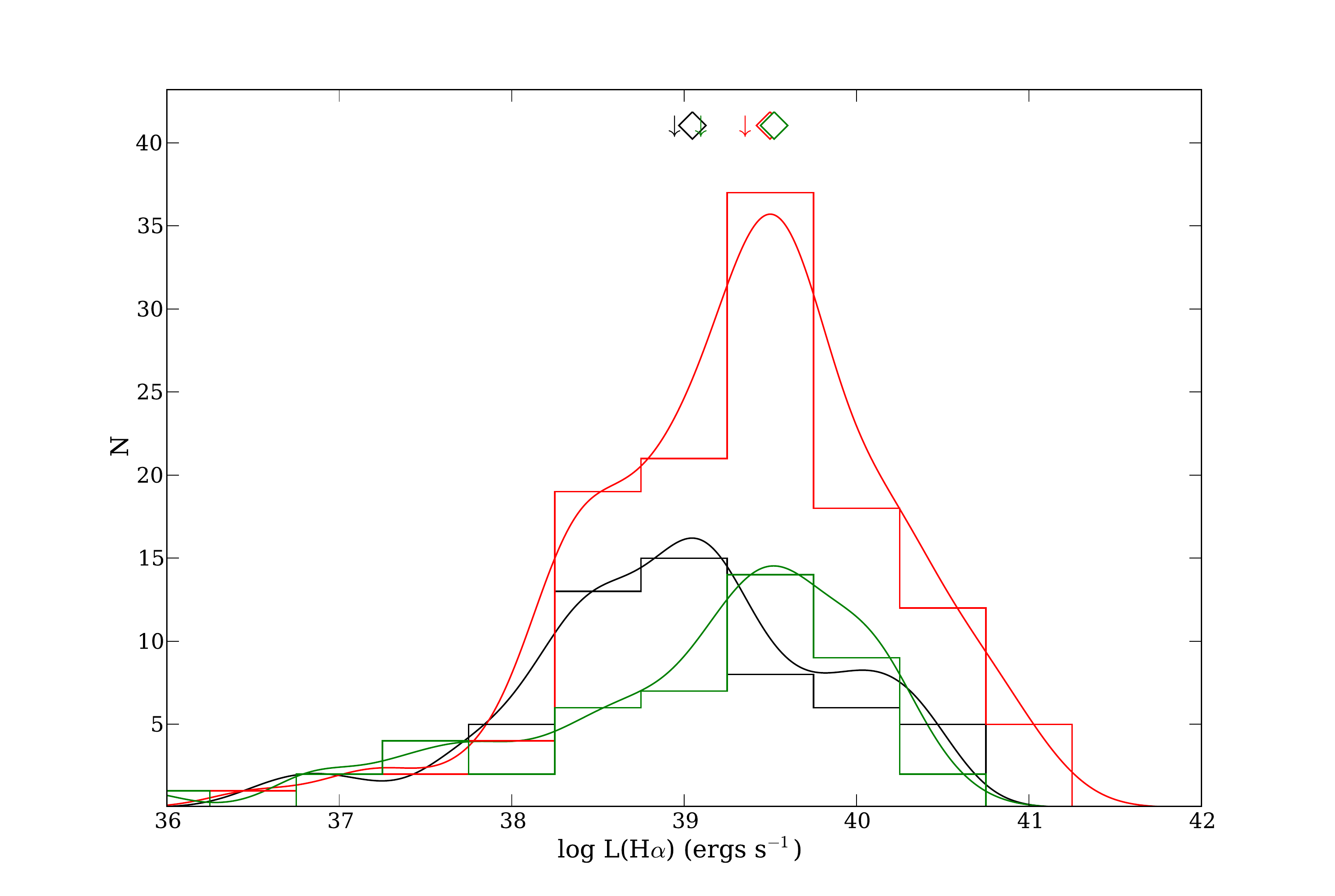}
\caption{\small Histogram of H$\alpha$ luminosities (regular and normalized
histograms).  The black curve is this sample, the red curve is the data
from Hunter \& Elmegreen for bright dwarf irregulars, the green
curve is the data from van Zee, a sample of HI-rich dwarf galaxies.  Arrows
mark the mean value for each sample, the diamonds mark the median.  Our
sample is noticeably fainter than Hunter \& Elmegreen sample primarily due
to the nature of their sample selection.  Our luminosities are similar to
van Zee's, with a slightly lower mean L(H$\alpha$).
}
\end{figure}

There is a strong correlation between absolute magnitude and L(H$\alpha$).
As can be seen in Figure 10, the correlation between stellar luminosity
(which measures stellar mass under an assumed IMF) and H$\alpha$
luminosity is evident in both the van Zee and Hunter\& Elmegreen samples
(red symbols) and our LSB sample (black symbols).  Neither H$\alpha$ flux
or apparent magnitude are correlated with distance, so we believe that
distance effects are not responsible for the correlation seen in Figure 10.
We note that there is a 0.5 shift in log L(H$\alpha$) between our samples
and the dwarf irregular samples (lower H$\alpha$ luminosities per stellar
mass for the LSB galaxies).  We have marked the eight galaxies in common
with Hunter \& Elmegreen, a bias flux calibration for our sample does not
seem evident as the explanation for this shift in the zeropoint.  There are
numerous possibilities for this difference and we will reserve speculation
until our analysis papers.

\begin{figure}[!t]
\centering
\includegraphics[bb=10 10 450 430,width=5.5in]{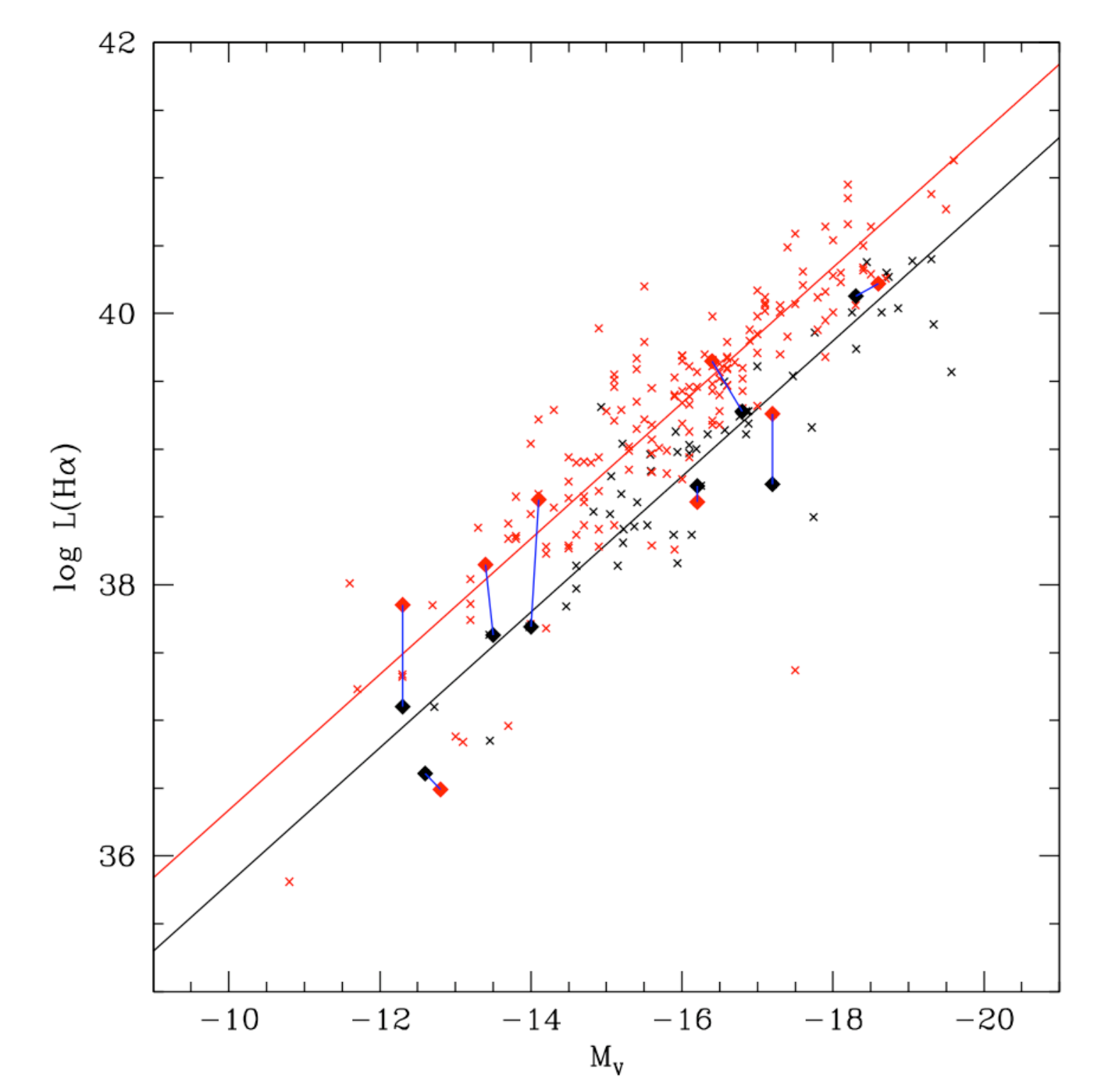}
\caption{\small Correlation between stellar mass (absolute magnitude) and
H$\alpha$ luminosity.  The black symbols are this paper's LSB sample.  The
red symbols are the combined samples of van Zee and Hunter \& Elmegreen
(there was no statisical difference between those two samples).  The eight
galaxies in common with Hunter \& Elmegreen are should as solid symbols
(connected for their differing H$\alpha$ values).  The LSB
galaxies are about 0.5 log L(H$\alpha$) fainter per stellar mass than the
dwarf irregular samples and there is no indication this is due to improper
H$\alpha$ fluxes.
}
\end{figure}

There is a weak correlation between central surface brightness ($\mu_c$)
and H$\alpha$ luminosity (see Figure 11) in the direction of weaker
L(H$\alpha$) for fainter surface brightnesses.  This is not detected in other
studies (e.g., van Zee 2000); however, it is not unexpected as there are
correlations between stellar density and HI surface density (de Blok,
McGaugh \& van der Hulst 1996).  This would indicate that stellar and gas
mass (roughly) follows star formation as evident by H$\alpha$ emission, a
local version of the Schmidt law (Kennicutt 1998); however, there are
numerous avenues of speculation from an increased number of ionizing UV
photons with higher stellar density to higher gas surface densities leading
to higher star formation rates.

\begin{figure}[!t]
\centering
\includegraphics[scale=0.7,angle=0]{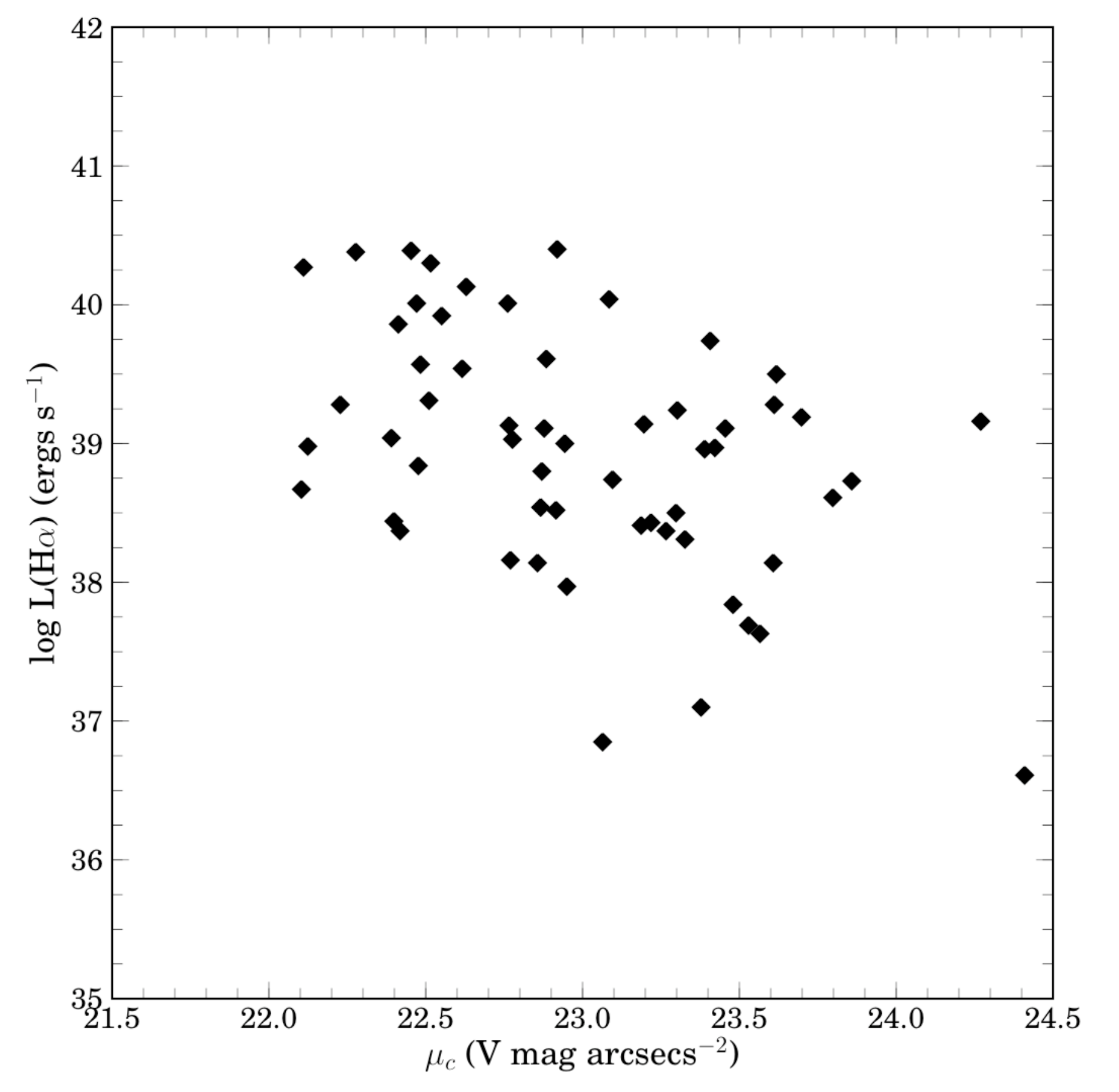}
\caption{\small Central surface brightness versus H$\alpha$ luminosity.
There is a weak correlation between central stellar density and the global
H$\alpha$ luminosity for LSB galaxies.  There are numerous avenues of speculation
from an increase number of ionizing UV 
photons with higher stellar density to higher gas surface densities leading 
to higher star formation rates.
}
\end{figure}

Three examples of the H$\alpha$ spatial distribution (D495-2, D570-3,
F611-1) are found in Figure 12 (all the images can be found at the data
website).  These three examples were selected primarily to display a common
feature to LSB galaxies, the fact that H$\alpha$ emission is frequently
uncorrelated with local stellar density.  While examples of stellar knots
associated with H$\alpha$ regions can be found (e.g., the brightest HII
region in D495-2 corresponds with a stellar knot), typically the HII knots
are not associated with any stellar enhancement (although there are
frequently blue `bubbles' in the $B-V$ color maps, see \S3.5).  These maps
should be compared with the H$\alpha$ image of UGC 128 (also found at the
data website), to fully display the difference H$\alpha$ morphology for
normal spirals and LSB's.

\subsection{Total Colors}

Total colors (and magnitudes) are extracted using elliptical apertures and
asymptotic fits to the apertures as a function of radius (Schombert 2007).
Mean colors, quoted in Table 1, are deduced from the color profiles where a
surface brightness average is taken between one scale length ($\alpha$) and
the $R_{V25}$ radius.

\begin{figure}[!t]
\centering
\includegraphics[scale=0.7,angle=0]{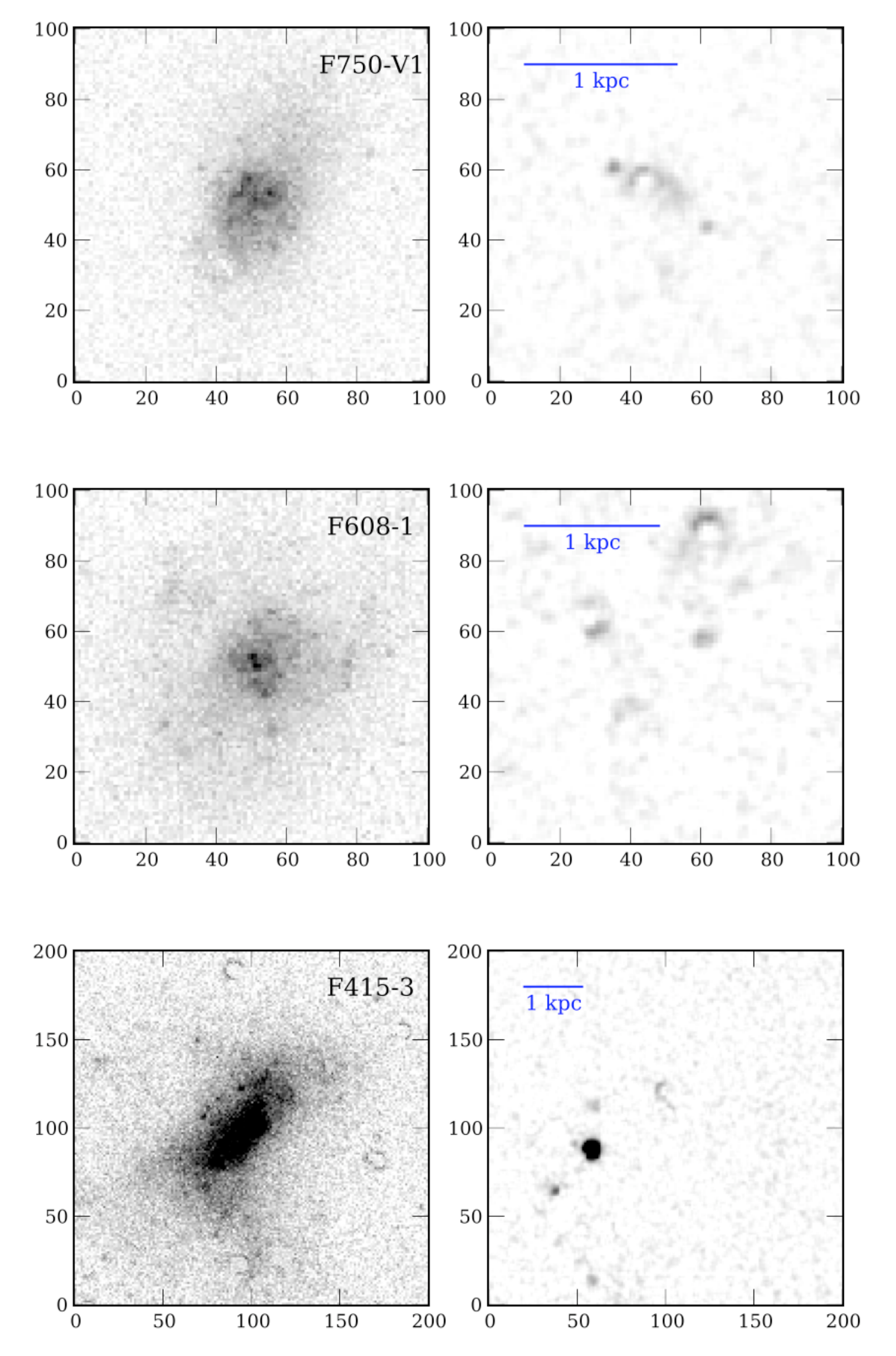}
\caption{\small Three examples of H$\alpha$ maps compared to $V$ images
($V$ on the left, H$\alpha$ on the right).
While on occasion HII regions are seen in the $V$ images (e.g., the brightest
knot in D495-2), in most cases the H$\alpha$ emission is not visible in the
optical images.  
}
\end{figure}

Our sample's distribution of mean $B-V$ color is shown in Figure 13.  Also
shown is the distribution of $B-V$ colors from the RC3 for galaxies with
morphological types later than Sc plus the combined samples from Hunter \&
Elmegreen (2006) and van Zee (2001) samples of star-forming dwarf irregulars.
Our sample has a similar mean and range as the other dwarf galaxy samples
(although we are not restricted to solely dwarf sized galaxies).  All three
dwarf samples are bluer than the late-type RC3 sample, probably reflecting
the known reddening in these dust rich, HSB systems.

There is no correlation between surface brightness and color, this has been
reported by other studies (McGaugh \& de Blok 1997, van Zee 2000).  There
was also no correlation observed in $V-I$ for the same sample (Pildis,
Schombert \& Eder 1997).  Other studies have larger ranges in central
surface brightness, and it can be seen in Figure 13 that HSB RC3 galaxies
are, on average, redder than LSB galaxies.  Our suspicion here is that a
combination of the limited range in surface brightness plus irregular
morphology masks any expected global trends with respect to color.

\begin{figure}[!t]
\centering
\includegraphics[scale=0.5,angle=0]{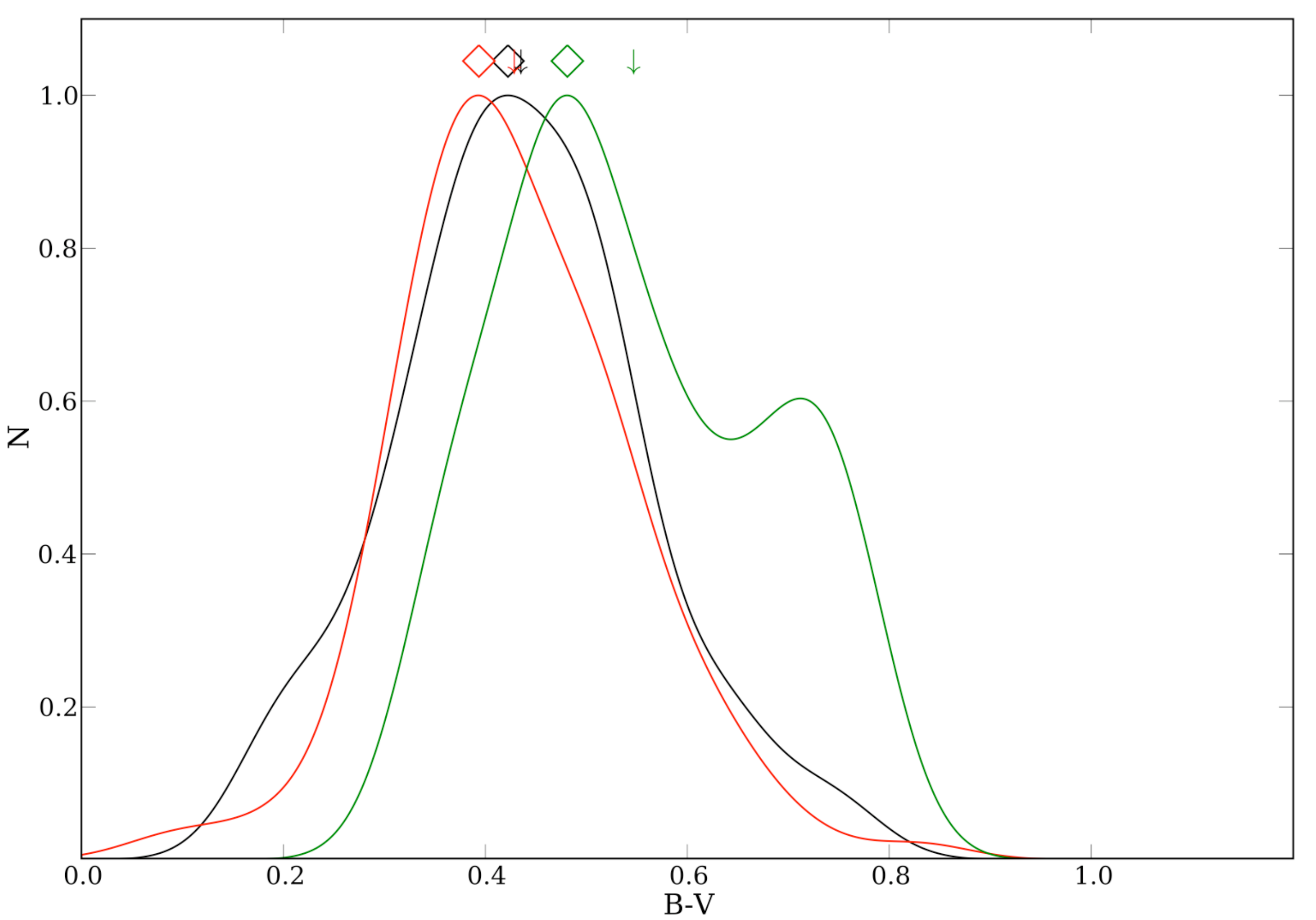}
\caption{\small Normalized histograms of $B-V$ color.  This sample is in
black, the RC3 late-type galaxies are in green, Hunter \& Elmegreen and van Zee dwarf
galaxies are in red.  Our sample has a similar mean and range as the dwarf
galaxy sample, both are significantly bluer than the RC3 sample.
}
\end{figure}

In addition to $B-V$ colors, 23 galaxies in our sample were also imaged in
the $I$ band (Pildis, Schombert \& Eder 1997).  The $B-V$ versus $V-I$
color plane is shown in Figure 14 along with $BVI$ data from the RC3.
While our LSB sample clearly occupies the bluest portion of the $BVI$
plane, it is also very much an extension of the color trend outlined by the
RC3 galaxies and overlaps with the bluest galaxies in the RC3 sample.
While there is some tendency for LSB galaxies to be bluer in $V-I$ compared
to the RC3 trend, the mean values are equal between the RC3 and our sample.

Also shown in Figure 14 are two stellar population models tracks
(instantaneous burst followed by passive evolution), one for a 13 Gyrs
composite stellar population (with a standard chemical evolutionary
scenario applied, see Schombert \& Rakos 2009) and a 1 Gyr population
([Fe/H] varies from the red to the blue tip by +0.4 to -2.0).  While this
is a simplistic comparison for galaxies which clearly have some (although
very little) current star formation, it does place LSB galaxies in the
context of galaxy evolution.  

The reddest RC3 galaxies are well explained by an old stellar population
that ranges in metallicity (Schombert \& Rakos 2009).  Given that the mean
metallicity of LSB galaxies is about [Fe/H] = -1.0 (McGaugh 1991), there is
no old stellar population that fits their $B-V$ versus $V-I$ colors.  This
indicates that there must be a significant younger population in LSB
galaxies, despite their low stellar density appearance.  This, of course,
is the core of the LSB dilemma, recent star formation to achieve blue
optical colors without large numbers of bright OB stars to increase its
luminosity density.  We will explore a larger range of star formation
models in a later paper.

\begin{figure}[!t]
\centering
\includegraphics[scale=0.7,angle=0]{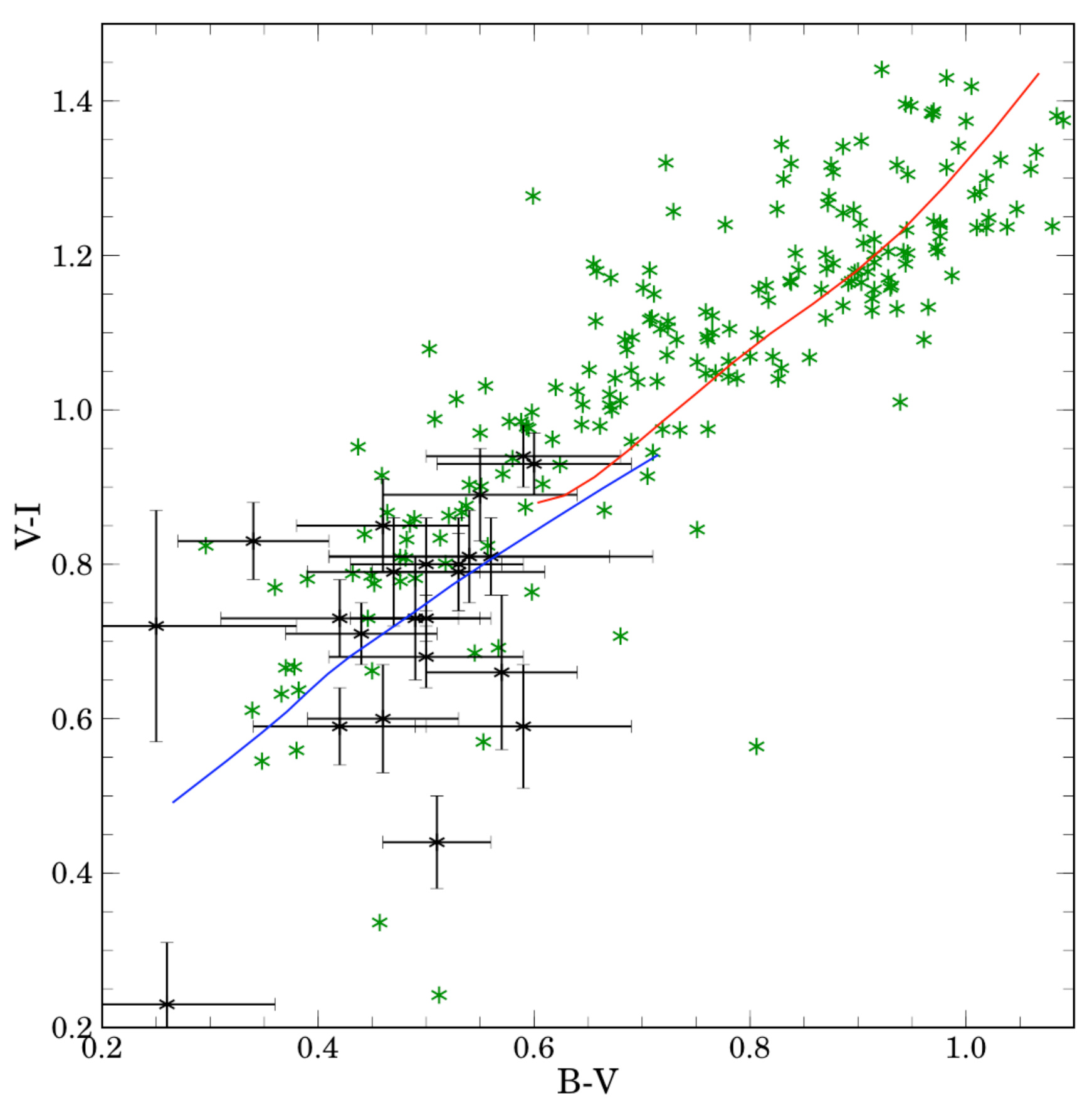}
\caption{\small $BVI$ color plane for the subsample of LSB galaxies with
$B-V$ and $V-I$ colors (black symbols).  RC3 galaxies are shown as green
symbols.  Also shown are 13 Gyrs stellar population tracks (instantaneous
burst followed by passive evolution) for a range of metallicities (red) and
a 1 Gyr stellar population (blue).  Given that the mean metallicity of LSB
galaxies is about [Fe/H] = -1.0 (McGaugh 1991), there is no old stellar
population that fits their $B-V$ versus $V-I$ colors.  Therefore, all
indicators are that there must be a significant younger population in LSB
galaxies, despite their low stellar density appearance.
}
\end{figure}

\subsection{$B-V$ Color Maps}

Deep $B$ and $V$ imaging allows, for the first time, a pixel-to-pixel
examination of the color distribution in LSB galaxies.  Three examples of
this type of color map is seen in the left panel of Figure 15, their
H$\alpha$ emission is shown in right panel.  There is no obvious
correlation between color and H$\alpha$ emission for these three galaxies.
However, this should not be too surprising as there is little evidence for
H$\alpha$ emission in the $V$ images (the H$\alpha$ emission is not visible
in stellar luminosity density).

We also note there are several notable exceptions to the lack of
color-H$\alpha$ correlation.  For example, there is a clear blue ridge in
the color map of D564-9 that corresponds to a linear H$\alpha$ feature
(we note that D564-9 is one of the most active star-forming galaxies in
this sample).  In addition, there are blue `star bubble' features
associated with HII regions in galaxies D631-7, D646-5 and D646-7.  Here
the blue colors are spatially correlated with the diffuse emission around a
HII region, rather than the core of the H$\alpha$ emission itself.  This
might reflect the evolutionary development of the stellar complex or it might
indicate that the stellar cores are obscured by dust (although LSB galaxies
are not prolific far-IR sources, Schombert \& Bothun 1988 and seem to
contain very little dust O'Neil \etal\ 1998).  These color features will be
explored in a later paper.

\begin{figure}[!t]
\centering
\includegraphics[scale=0.6,angle=0]{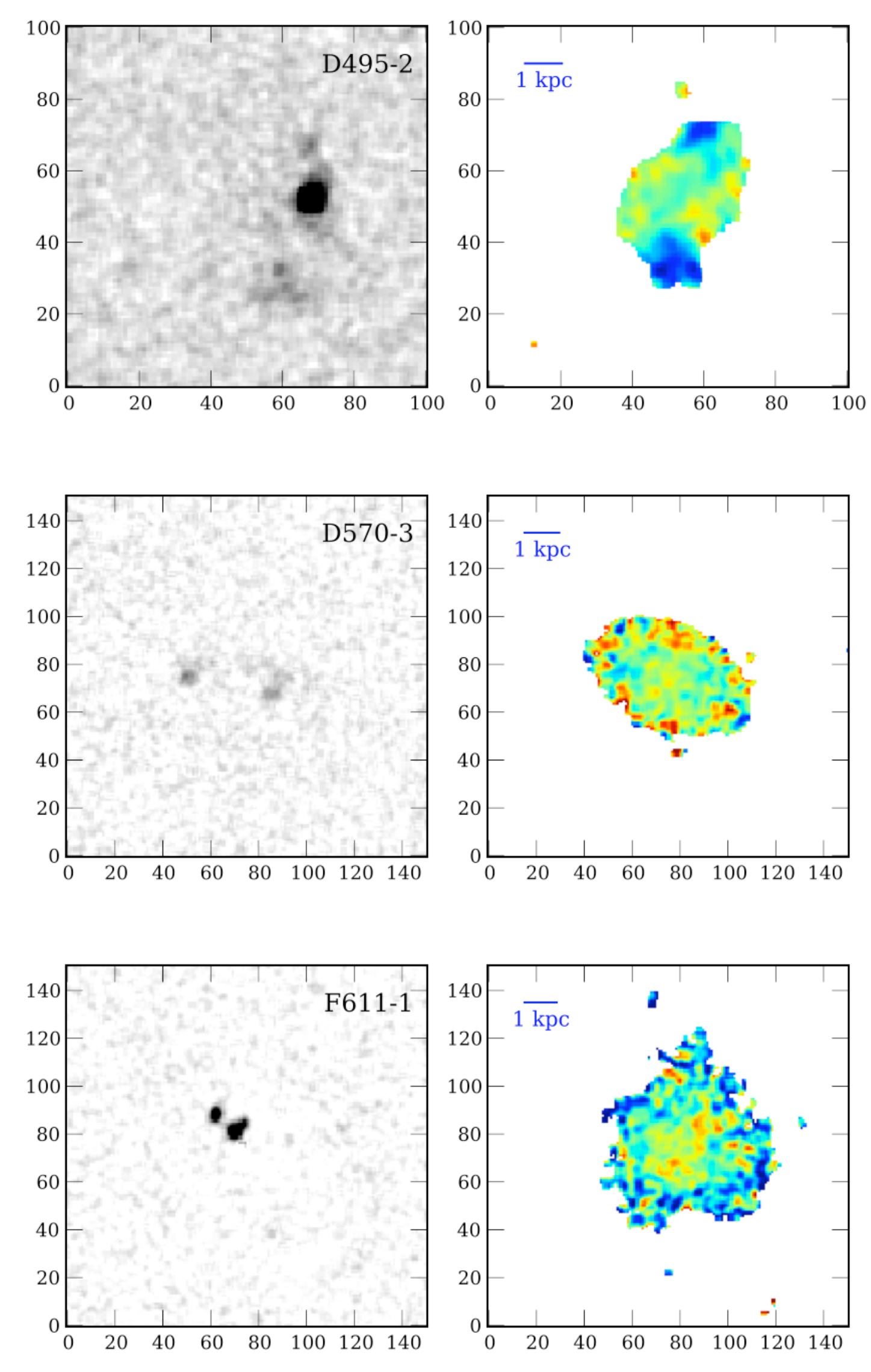}
\caption{\small Three LSB galaxies viewed for their H$\alpha$ emission
(left panel) and their $B-V$ color maps (right panel, $B-V = 0.0$ is blue,
$B-V = 1.0$ is red).  There are no prominent correlates with spatial color
and H$\alpha$ emission.
}
\end{figure}

This dataset is also sufficiently high in S/N to allow us to compare the
color of individual pixels with the pixel's surface brightness.  This is
accomplished by assigning to each pixel a mean $B-V$ color and a surface
brightness, based on its calibrated $V$ flux divided by pixel area. This
plot, using 280,000 pixels above 25 $V$ mag arcsecs$^{-2}$, is shown in top
panel of Figure 16.  Each color-$\mu$ data point is treated as a 2D
gaussian with a standard deviation tied to the color and surface brightness
error of the pixel.  All the pixels are summed and binned to produce the
density diagrams in Figure 16.

The bottom panel displays the same colors for but only for pixels with
detectable H$\alpha$ emission.  Note that the brightest pixels (above 23.5
mag arcsecs$^{-2}$) are the bluest pixels in either plot.  H$\alpha$ pixels
can occur in regions of high and low surface brightness, with the same
distribution of color as other pixels.  And, while H$\alpha$ pixels are
associated with some high surface brightness areas, the reverse is not true
and, therefore, it is difficult to isolate star formation regions just from
$V$ band images as noted above.

The increasing spread of color with fainter surface brightness is reflecting
increasing error at low S/N levels.  Even with this spread there is a weak
correlation between pixel color and surface brightness such that the
brightest pixels are the bluest.  While this is not surprising, and also
true of normal spirals, it does, at least, tie star formation in LSB
galaxies as a similar process to that in normal spirals (i.e. the bluest
regions are associated with bright, high mass stars, not just lowest
metallicities).

\begin{figure}[!t]
\centering
\includegraphics[scale=0.6,angle=0]{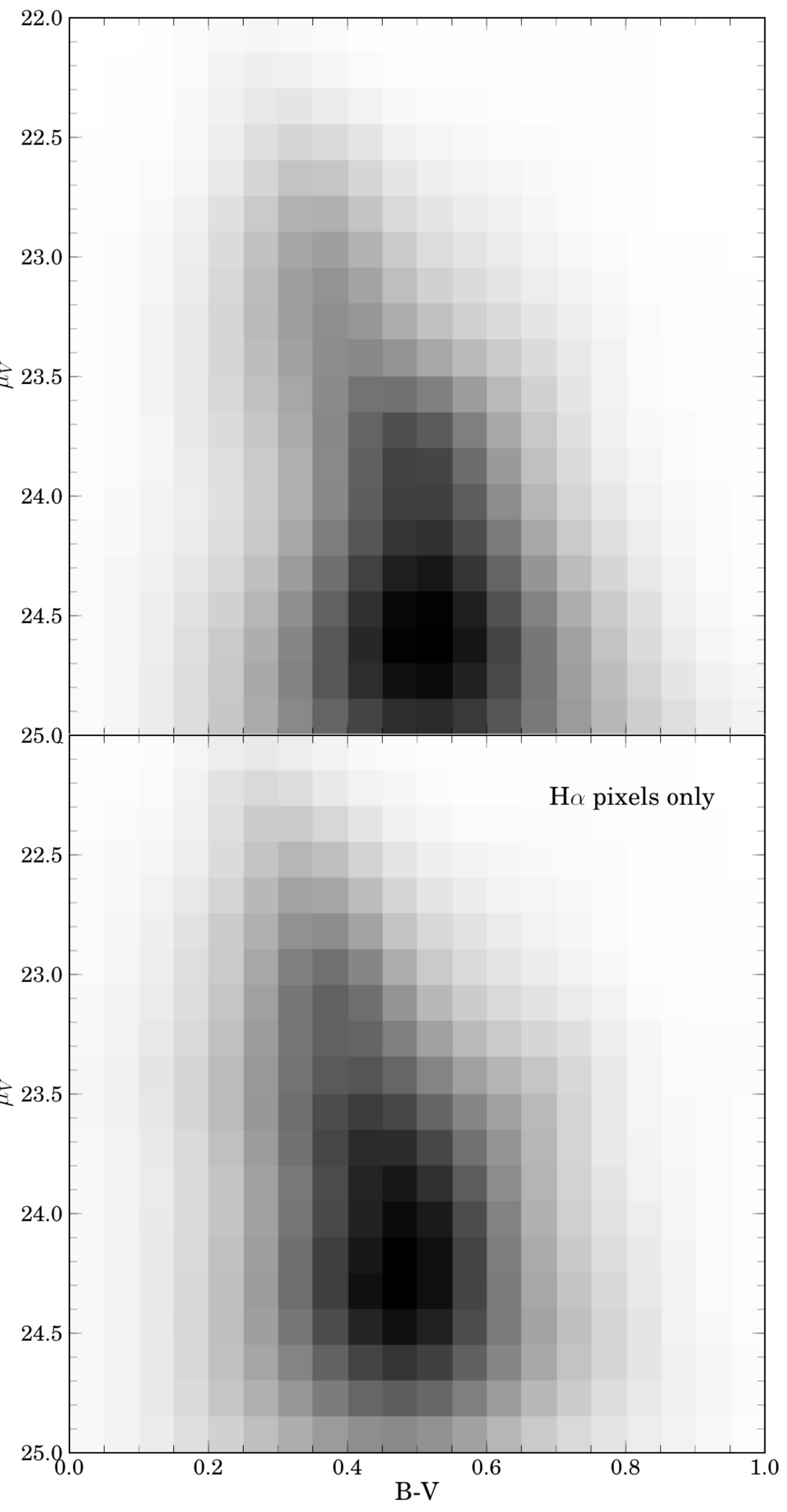}
\caption{\small The pixel-to-pixel color-surface brightness map for all 58
galaxies in the sample.  A total of 280,000 pixels are summed, then binned
to produce the density plot of $B-V$ color versus surface brightness on
pixel scales.  The top panel displays all pixels above 25 $V$ mag
arcsecs$^{-2}$, the bottom panel only those pixels associated with
H$\alpha$ emission.
}
\end{figure}

The same analysis can be performed on the $V-I$ color images for a
subsample of galaxies with $BVI$ photometry.  The two color pixel diagram
is shown in Figure 17.  The only clear difference between the pixel colors
and the global galaxy colors in Figure 14 is a distinct tail of bluer $V-I$
colors at $B-V = 0.6$.  These are fainter surface brightness pixels, but
their origin is unclear as none of the simpler stellar population models
occupy this portion of the $BVI$ plane for any combination of age and
metallicity.  We will discuss this problem in a later paper.

\begin{figure}[!t]
\centering
\includegraphics[scale=0.8,angle=0]{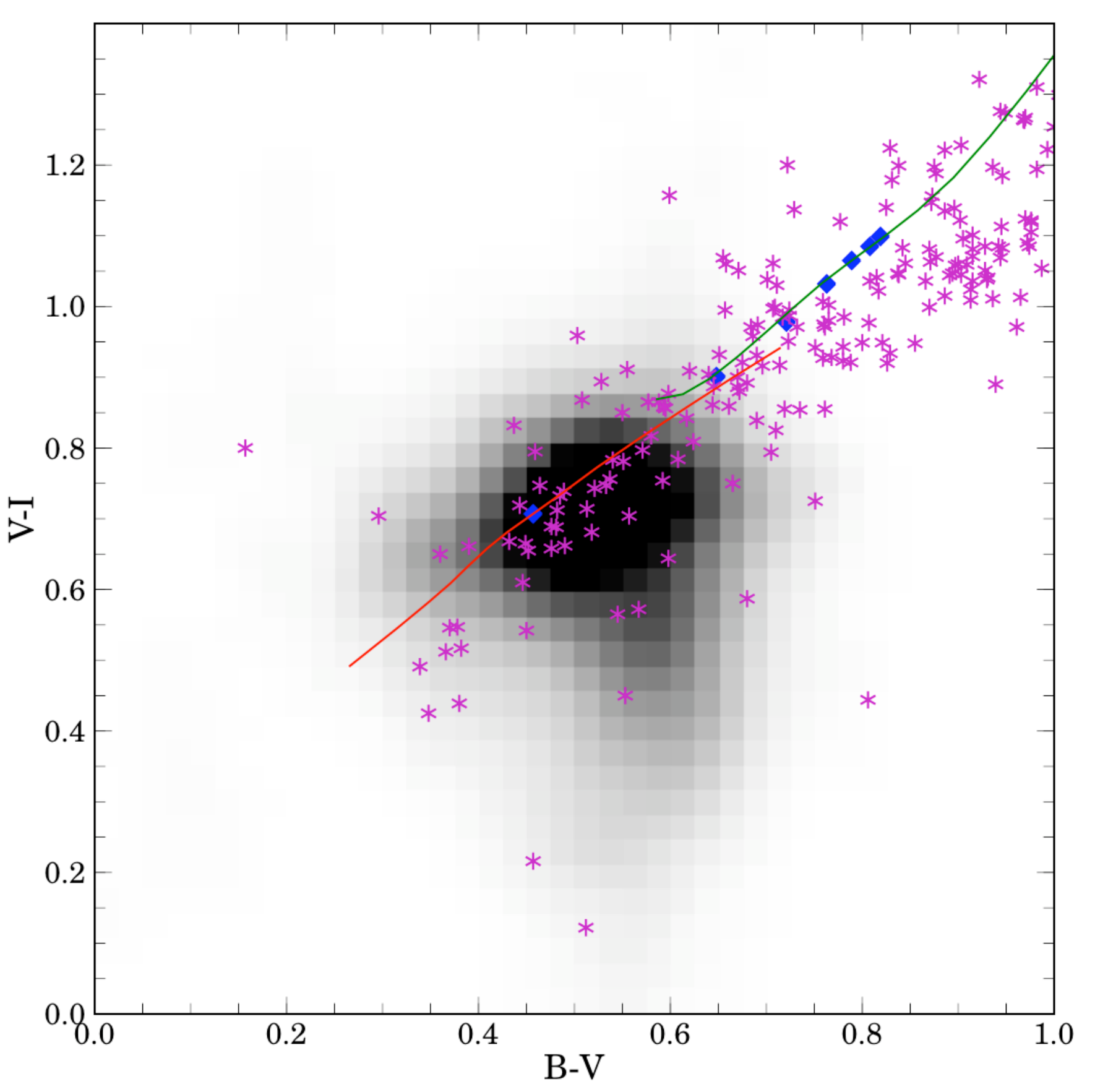}
\caption{\small $BVI$ color plane for pixels in LSB galaxies.  Pixels were
selected to be above 25 $V$ mag arcsecs$^{-2}$, solid symbols are again RC3
galaxies plus 13 and 1 Gyrs stellar population tracks.
}
\end{figure}

\section{Summary} 

We present optical and H$\alpha$ imaging for a sample of 61 LSB galaxies
selected from the PSS-II LSB catalog (Schombert \etal 1992).  Combined with
previous HI measurements, this provides data for a comprehensive study of the
gas fraction, current star formation rates, stellar population characteristics and
mode of star formation.  This paper presents only the core data.  Analysis
on the meaning of the data will be offered in the later papers of this
series.  A summary of the characteristics of the dataset are:

\begin{description}

\item{(1)} The sample of LSB galaxies selected for our study was based on
their morphology and redshift (less than 8,000 km/sec for the H$\alpha$
filter set).  No selection was made on luminosity, size or dynamical mass.
The sample covers a range of galaxy characteristics that overlap with
magnitude selected samples (e.g., RC3) and dwarf galaxy samples (van Zee
2000).

\item{(2)} Data analysis attempts to present the sample in both a visual
way (see Figure 1) and using an XML data format that allows the user to
follow the calibration and reduction process.

\item{(3)} Total magnitudes and colors are extracted using asymptotic fits
and surface brightness averaged color isophotes.  The resulting values are
in agreement, within the errors, with galaxies in common from the Hunter \&
Elmegreen (2004) sample of dwarf irregulars.

\item{(4)} Central surface brightness and isophotal size are extracted from
surface brightness profiles of each galaxy ($\mu_c$ and R$_{V25}$).
Exponential fits are adequate for about 80\% of the sample, providing disk
scale length ($\alpha$ and central surface brightness ($\mu_o$).

\item{(5)} H$\alpha$ luminosities are extracted by grid photometry.
Comparison to aperture values indicates that a 40\% correction to account
for diffuse emission is required (in agreement with van Zee 2000).
Resulting L(H$\alpha$) values are in agreement, within the errors, with
Hunter \& Elmegreen.  Our sample has a mean H$\alpha$ luminosity about a
factor of 10 lower than other dwarf or gas-rich galaxy studies (e.g. van
Zee 2000).

\item{(6)} H$\alpha$ maps confirm the sporadic nature of star formation in
LSB galaxies.  The HII regions are neither coherent to themselves (i.e. a
density wave) or any optical features in the galaxy.  The only correlation
between a global property of an LSB and its star formation rate is total
stellar mass and L(H$\alpha$).

\item{(7)} LSB color distributions in $B-V$ and $V-I$ are similar to dwarf
irregular samples, but about 0.1 bluer in $B-V$ than late-type spirals.
Two color diagrams place LSB galaxies at the extreme blue end of the color
distribution of all galaxy types.  

\item{(8)} Color maps display little correlation between color and
H$\alpha$ emission.  However, several color features are found to associate
with some HII regions (e.g. blue shells).  Color gradients vary in
direction and magnitude for the sample.

\end{description}

In general, this dataset presents an overall picture of LSB galaxies where
they have some characteristics in common with dwarf galaxies and gas-rich
irregulars, yet differ in significant ways that may provide the clue to
their LSB nature.  They cover a range of sizes and luminosities beyond the
definition of dwarf galaxies, yet have SFR's lower than galaxies of the
comparable luminosity (i.e. stellar mass).  In the later papers of this
series, we will present an analysis of the complete dataset with the goal
of outlining possible star formation histories for LSB galaxies.

\acknowledgements We gratefully acknowledge KPNO/NOAO for the telescope
time to complete this project.  Software for this project was developed
under NASA's AIRS and ADP Programs.

\clearpage

\begin{deluxetable}{lcccccccc}
\tablecolumns{8}
\small
\tablewidth{0pt}
\tablecaption{Optical Properties}
\tablehead{
\colhead{Object} & \colhead{run} & \colhead{D} & \colhead{$M_V$} & \colhead{$\mu_o$} & 
\colhead{$\alpha$} & \colhead{$<B-V>$} & \colhead{$<V-I>$} & \colhead{$(b/a)_{25}$} \\
\colhead{} & \colhead{} & \colhead{(Mpc)} & \colhead{} & \colhead{(V mag arcsec$^{-2}$)} &
\colhead{(kpc)} & \colhead{} & \colhead{} & \colhead{} \\
}
\startdata

D495-1     &  K0308 &  34.9 & -15.94$\pm$0.02 & 22.29 & 0.8 & 0.50 & 0.72 & 0.56 \\
D495-2     &  K0308 &  33.4 & -15.92$\pm$0.03 & 21.90 & 0.6 & 0.42 & 0.58 & 0.58 \\
D500-3     &  K0308 &  22.7 & -15.59$\pm$0.04 & 21.04 & 0.4 & 0.48 & 0.28 & 0.69 \\
D563-1     &  K0309 &  61.6 & -16.85$\pm$0.06 & 22.68 & 1.1 & 0.44 & 0.69 & 0.81 \\
D564-9     &  K0308 &  46.0 & -18.45$\pm$0.04 & 22.18 & 2.1 & 0.40 & 0.59 & 0.65 \\
D564-16    &  K0308 & 100.0 & -17.74$\pm$0.09 & 23.00 & 2.4 & 0.39 &  --  & 0.50 \\
D565-10    &  K0309 &  11.8 & -14.33$\pm$0.03 & 22.63 & 0.3 & 0.73 & 0.77 & 0.91 \\
D568-2     &  K0309 &  21.3 & -14.60$\pm$0.02 & 21.67 & 0.3 & 0.42 & 0.66 & 0.50 \\
D570-3     &  K0308 &  23.7 & -15.23$\pm$0.02 & 22.61 & 0.5 & 0.56 & 0.53 & 0.73 \\
D570-7     &  K0308 &  15.6 & -14.47$\pm$0.07 & 22.99 & 0.5 & 0.44 &  --  & 0.48 \\
D572-2     &  K0308 &  56.5 & -16.34$\pm$0.17 & 22.16 & 1.4 & 0.20 & 0.48 & 0.24 \\
D572-5     &  K0309 &  18.0 & -15.20$\pm$0.02 & 21.51 & 0.3 & 0.40 & 0.48 & 0.69 \\
D575-2     &  K0308 &  14.7 & -16.10$\pm$0.05 & 23.37 & 1.4 & 0.52 & 0.80 & 0.52 \\
D575-7     &  K0309 &  18.1 & -15.06$\pm$0.02 & 21.73 & 0.4 & 0.40 & 0.61 & 0.59 \\
D631-7     &  K0308 &   7.8 & -15.21$\pm$0.03 & 21.27 & 0.4 & 0.38 &  --  & 0.45 \\
D637-3     &  K0308 &  35.2 & -15.58$\pm$0.15 & 23.27 & 1.0 & 0.34 &  --  & 0.61 \\
D646-5     &  K0308 &  18.3 & -14.83$\pm$0.05 & 23.01 & 0.6 & 0.37 & 1.81 & 0.56 \\
D646-7     &  K0308 &   7.2 & -14.93$\pm$0.03 & 22.43 & 0.4 & 0.39 & 0.73 & 0.66 \\
D646-8     &  K0308 &  33.3 & -14.71$\pm$0.40 & 24.07 & 1.0 & 0.49 &  --  & 0.67 \\
D646-9     &  K0308 &  29.4 & -15.15$\pm$0.14 & 23.57 & 0.9 & 0.32 &  --  & 0.59 \\
D646-11    &  K0309 &  12.1 & -15.05$\pm$0.03 & 22.50 & 0.5 & 0.52 & 0.44 & 0.74 \\
D656-2     &  K0309 &  15.9 & -15.37$\pm$0.12 & 23.09 & 0.7 & 0.33 & 0.55 & 0.77 \\
D723-4     &  K0309 &  32.9 & -16.19$\pm$0.10 & 22.64 & 1.0 & 0.41 & 0.64 & 0.51 \\
D723-5     &  K0309 &  27.7 & -16.88$\pm$0.02 & 21.97 & 0.9 & 0.53 & 0.72 & 0.54 \\
D723-9     &  K0308 &  26.2 & -17.47$\pm$0.03 & 22.88 & 2.2 & 0.63 & 0.69 & 0.37 \\
D774-1     &  K0309 &  72.0 & -19.57$\pm$0.01 & 22.29 & 2.2 & 0.52 & 0.61 & 0.74 \\
DDO154     &  K0309 &   8.9 & -16.10$\pm$0.04 & 21.73 & 0.6 & 0.31 &  --  & 0.45 \\
DDO168     &  K0309 &   5.2 & -15.94$\pm$0.12 & 21.06 & 0.5 & 0.35 &  --  & 0.36 \\
F415-3     &  K1007 &  10.4 & -15.22$\pm$0.01 & 22.84 & 0.7 & 0.52 &  --  & 0.31 \\
F473-V2    &  K1007 &  44.7 & -16.13$\pm$0.15 & 23.12 & 1.5 & 0.44 &  --  & 0.34 \\
F512-1     &  K0309 &  14.1 & -15.54$\pm$0.07 & 22.40 & 0.6 & 0.19 & 0.79 & 0.55 \\
F533-1     &  K1007 &  12.8 & -13.98$\pm$0.08 & 23.24 & 0.4 & 0.29 &  --  & 0.76 \\
F544-1     &  K1007 &  28.5 & -15.89$\pm$0.08 & 22.52 & 0.9 & 0.26 &  --  & 0.60 \\
F561-1     &  K0308 &  69.8 & -18.30$\pm$0.02 & 22.47 & 1.6 & 0.64 &  --  & 1.00 \\
F562-V1    &  K0308 &  68.1 & -18.25$\pm$0.06 & 21.85 & 1.8 & 0.47 & 0.68 & 0.19 \\
F563-1     &  K0308 &  52.2 & -17.76$\pm$0.05 & 22.36 & 1.5 & 0.37 & 0.82 & 0.65 \\
F563-V1    &  K0308 &  57.6 & -17.20$\pm$0.06 & 23.08 & 2.0 & 0.20 & 0.79 & 0.47 \\
F564-V3    &  K0309 &  10.4 & -14.05$\pm$0.04 & 23.43 & 0.4 & 0.64 & 0.88 & 0.81 \\
F565-V1    &  K0308 &  10.8 & -13.46$\pm$0.08 & 22.84 & 0.3 & 0.21 &  --  & 0.70 \\
F565-V2    &  K0309 &  55.1 & -16.25$\pm$0.32 & 23.42 & 1.6 & 0.41 &  --  & 0.50 \\
F568-1     &  K0308 &  95.5 & -18.86$\pm$0.08 & 22.93 & 3.1 & 0.50 &  --  & 0.67 \\
F568-V1    &  K0309 &  84.8 & -18.64$\pm$0.03 & 22.56 & 2.3 & 0.44 &  --  & 0.68 \\
F574-1     &  K0309 & 100.0 & -19.05$\pm$0.11 & 22.12 & 3.2 & 0.48 &  --  & 0.36 \\
F574-2     &  K0309 &  92.3 & -18.31$\pm$0.16 & 23.49 & 4.7 & 0.54 &  --  & 0.38 \\
F575-3     &  K0308 &   9.8 & -14.31$\pm$0.09 & 23.80 & 0.8 & 0.26 &  --  & 0.30 \\
F577-V1    &  K0309 & 113.0 & -18.74$\pm$0.06 & 21.98 & 2.0 & 0.48 &  --  & 0.20 \\
F579-V1    &  K0309 &  90.5 & -19.33$\pm$0.02 & 21.94 & 2.1 & 0.64 &  --  & 0.95 \\
F583-2     &  K0309 &  25.4 & -16.57$\pm$0.11 & 23.10 & 1.4 & 0.50 &  --  & 0.55 \\
F608-1     &  K1007 &   9.0 & -13.45$\pm$0.05 & 23.74 & 0.4 & 0.50 &  --  & 0.86 \\
F608-V1    &  K1007 &  20.3 & -14.60$\pm$0.12 & 22.47 & 0.4 & 0.38 &  --  & 0.72 \\
F611-1     &  K1007 &  25.5 & -15.41$\pm$0.03 & 22.97 & 0.7 & 0.48 &  --  & 0.75 \\
F612-V3    &  K1007 &  65.4 & -16.56$\pm$0.09 & 22.80 & 1.3 & 0.41 &  --  & 0.34 \\
F614-V2    &  K1007 &  51.3 & -17.00$\pm$0.05 & 21.98 & 1.2 & 0.40 &  --  & 0.60 \\
F615-1     &  K1007 &   8.2 & -12.60$\pm$0.06 & 23.14 & 0.3 & 0.34 &  --  & 0.45 \\
F651-2     &  K0308 &  27.5 & -16.85$\pm$0.07 & 23.10 & 1.8 & 0.55 &  --  & 0.47 \\
F677-V2    &  K1007 &  63.9 & -16.76$\pm$0.04 & 22.84 & 1.2 & 0.75 &  --  & 0.88 \\
F687-1     &  K1007 &  47.5 & -17.72$\pm$0.04 & 23.98 & 3.9 & 0.48 &  --  & 0.64 \\
F750-2     &  K1007 &  46.1 & -16.88$\pm$0.07 & 22.30 & 1.4 & 0.40 &  --  & 0.28 \\
F750-V1    &  K1007 &   8.0 & -12.72$\pm$0.04 & 22.68 & 0.2 & 0.32 &  --  & 0.74 \\
U128       &  K1007 &  58.5 & -19.30$\pm$0.03 & 22.60 & 3.6 & 0.57 &  --  & 0.53 \\
U5005      &  K0309 &  57.1 & -18.71$\pm$0.01 & 22.36 & 2.2 & 0.32 &  --  & 0.38 \\

\enddata
\end{deluxetable}

\begin{deluxetable}{lccccccc}
\tablecolumns{8}
\small
\tablewidth{0pt}
\tablecaption{SF and Gas Properties}
\tablehead{
\colhead{Object} &
\colhead{log $f_{H\alpha}$} &
\colhead{log $L_{H\alpha}$} &
\colhead{log $M_*$} &
\colhead{log $M_{HI}$} &
\colhead{log $M_{b}$} &
\colhead{$f_g$} &
\colhead{b} \\

\colhead{} &
\colhead{(ergs s$^{-1}$ cm$^{-2}$)} &
\colhead{(ergs s$^{-1}$)} &
\colhead{($M_{\sun}$)} &
\colhead{($M_{\sun}$)} &
\colhead{($M_{\sun}$)} &
\colhead{} &
\colhead{} \\

}
\startdata

D495-1     & -15.03$\pm$0.08 & 38.16 &  7.93 &  8.23 &  8.40 &  0.66 & -0.87 \\
D495-2     & -14.02$\pm$0.04 & 39.13 &  7.81 &  8.42 &  8.52 &  0.80 &  0.21 \\
D500-3     & -13.96$\pm$0.10 & 38.84 &  7.76 &  8.28 &  8.40 &  0.77 & -0.02 \\
D563-1     & -14.57$\pm$0.09 & 39.11 &  8.21 &  8.85 &  8.94 &  0.81 & -0.20 \\
D564-9     & -13.05$\pm$0.02 & 40.38 &  8.80 &  9.59 &  9.66 &  0.86 &  0.47 \\
D564-16    & -15.59$\pm$1.50 & 38.50 &  8.51 &  8.93 &  9.07 &  0.72 & -1.11 \\
D565-10    &        none     & none  &  7.58 &  7.56 &  7.87 &  0.49 &  0.00 \\
D568-2     & -14.60$\pm$0.03 & 38.14 &  7.29 &  7.98 &  8.06 &  0.83 & -0.25 \\
D570-3     & -14.42$\pm$0.12 & 38.41 &  7.72 &  7.96 &  8.16 &  0.64 & -0.41 \\
D570-7     & -14.62$\pm$0.04 & 37.84 &  7.26 &  7.97 &  8.04 &  0.84 & -0.52 \\
D572-2     & -14.48$\pm$0.05 & 39.11 &  7.71 &  8.91 &  8.94 &  0.94 &  0.30 \\
D572-5     & -13.92$\pm$0.02 & 38.67 &  7.51 &  8.40 &  8.46 &  0.89 &  0.06 \\
D575-2     & -13.45$\pm$0.09 & 38.97 &  8.02 &  8.81 &  8.87 &  0.86 & -0.16 \\
D575-7     & -13.79$\pm$0.05 & 38.80 &  7.45 &  8.46 &  8.50 &  0.91 &  0.25 \\
D631-7     & -12.88$\pm$0.07 & 39.04 &  7.49 &  8.52 &  8.55 &  0.91 &  0.45 \\
D637-3     & -14.23$\pm$0.10 & 38.96 &  7.59 &  8.83 &  8.85 &  0.95 &  0.27 \\
D646-5     & -14.06$\pm$0.16 & 38.54 &  7.32 &  8.47 &  8.50 &  0.93 &  0.12 \\
D646-7     & -12.48$\pm$0.01 & 39.31 &  7.39 &  8.10 &  8.18 &  0.84 &  0.82 \\
D646-8     &        none     & none  &  7.42 &  7.67 &  7.87 &  0.64 &  0.00 \\
D646-9     & -14.88$\pm$0.05 & 38.14 &  7.39 &  8.65 &  8.67 &  0.95 & -0.35 \\
D646-11    & -13.73$\pm$0.06 & 38.52 &  7.60 &  8.02 &  8.16 &  0.72 & -0.18 \\
D656-2     & -14.06$\pm$0.11 & 38.43 &  7.49 &  8.57 &  8.60 &  0.92 & -0.15 \\
D723-4     & -14.11$\pm$0.07 & 39.00 &  7.91 &  9.13 &  9.15 &  0.94 & -0.01 \\
D723-5     & -13.69$\pm$0.03 & 39.28 &  8.34 &  8.46 &  8.70 &  0.56 & -0.17 \\
D723-9     & -13.38$\pm$0.02 & 39.54 &  8.71 &  9.02 &  9.19 &  0.67 & -0.27 \\
D774-1     & -14.31$\pm$0.08 & 39.57 &  9.40 &  9.70 &  9.88 &  0.66 & -0.93 \\
DDO154     & -12.95$\pm$0.02 & 39.03 &  7.75 &  9.31 &  9.32 &  0.97 &  0.18 \\
DDO168     & -12.54$\pm$0.02 & 38.98 &  7.75 &  8.81 &  8.85 &  0.92 &  0.13 \\
F415-3     & -13.71$\pm$0.09 & 38.31 &  7.67 &  8.68 &  8.72 &  0.91 & -0.46 \\
F473-V2    & -14.92$\pm$0.78 & 38.37 &  7.93 &  8.98 &  9.02 &  0.92 & -0.65 \\
F512-1     & -13.94$\pm$0.03 & 38.44 &  7.37 &  8.23 &  8.28 &  0.88 & -0.04 \\
F533-1     & -14.50$\pm$0.04 & 37.69 &  6.87 &  8.14 &  8.16 &  0.95 & -0.29 \\
F544-1     & -14.54$\pm$0.02 & 38.37 &  7.60 &  9.11 &  9.12 &  0.97 & -0.33 \\
F561-1     & -13.67$\pm$0.06 & 40.13 &  9.06 &  9.29 &  9.49 &  0.63 & -0.03 \\
F562-V1    & -13.76$\pm$0.03 & 40.01 &  8.82 &  9.53 &  9.61 &  0.84 &  0.09 \\
F563-1     & -13.68$\pm$0.04 & 39.86 &  8.49 &  9.69 &  9.72 &  0.94 &  0.27 \\
F563-V1    & -14.89$\pm$0.12 & 38.74 &  8.05 &  8.97 &  9.02 &  0.89 & -0.42 \\
F564-V3    &        none     & none  &  7.36 &  7.77 &  7.91 &  0.72 &  0.00 \\
F565-V1    & -15.31$\pm$0.30 & 36.85 &  6.57 &  7.03 &  7.16 &  0.74 & -0.82 \\
F565-V2    & -14.85$\pm$0.13 & 38.73 &  7.94 &  9.02 &  9.05 &  0.92 & -0.32 \\
F568-1     & -14.00$\pm$0.10 & 40.04 &  9.10 &  9.75 &  9.84 &  0.82 & -0.16 \\
F568-V1    & -13.93$\pm$0.02 & 40.01 &  8.94 &  9.74 &  9.80 &  0.86 & -0.02 \\
F574-1     & -13.69$\pm$0.02 & 40.39 &  9.16 &  9.76 &  9.86 &  0.80 &  0.14 \\
F574-2     & -14.27$\pm$0.06 & 39.74 &  8.93 &  9.32 &  9.47 &  0.71 & -0.30 \\
F575-3     &        none     & none  &  6.97 &  8.07 &  8.11 &  0.93 &  0.00 \\
F577-V1    & -13.92$\pm$0.05 & 40.27 &  9.02 &  9.81 &  9.87 &  0.86 &  0.14 \\
F579-V1    & -14.08$\pm$0.06 & 39.92 &  9.47 &  9.52 &  9.79 &  0.53 & -0.65 \\
F583-2     & -13.76$\pm$0.04 & 39.14 &  8.18 &  9.02 &  9.08 &  0.88 & -0.14 \\
F608-1     & -14.27$\pm$0.06 & 37.63 &  6.93 &  7.76 &  7.82 &  0.87 & -0.40 \\
F608-V1    & -14.65$\pm$0.06 & 37.97 &  7.24 &  8.59 &  8.61 &  0.96 & -0.37 \\
F611-1     & -14.20$\pm$0.01 & 38.61 &  7.69 &  8.55 &  8.60 &  0.88 & -0.18 \\
F612-V3    & -14.13$\pm$0.02 & 39.50 &  8.07 &  9.09 &  9.13 &  0.91 &  0.34 \\
F614-V2    & -13.81$\pm$0.04 & 39.61 &  8.23 &  9.09 &  9.14 &  0.88 &  0.28 \\
F615-1     & -15.21$\pm$0.08 & 36.61 &  6.39 &  7.58 &  7.61 &  0.94 & -0.89 \\
F651-2     & -13.69$\pm$0.07 & 39.28 &  8.36 &  9.03 &  9.12 &  0.83 & -0.18 \\
F677-V2    & -14.37$\pm$0.03 & 39.24 &  8.58 &  8.88 &  9.05 &  0.67 & -0.44 \\
F687-1     & -14.20$\pm$0.04 & 39.16 &  8.62 &  9.35 &  9.42 &  0.84 & -0.56 \\
F750-2     & -14.15$\pm$0.07 & 39.19 &  8.18 &  9.31 &  9.34 &  0.93 & -0.10 \\
F750-V1    & -14.71$\pm$0.07 & 37.10 &  6.40 &  7.17 &  7.24 &  0.85 & -0.41 \\
U128       & -13.11$\pm$0.02 & 40.40 &  9.36 & 10.16 & 10.22 &  0.86 & -0.06 \\
U5005      & -13.31$\pm$0.06 & 40.30 &  8.80 &  9.87 &  9.91 &  0.92 &  0.39 \\

\enddata
\end{deluxetable}

\end{document}